\documentclass[twoside,11pt]{article}
\usepackage{a4wide,cite,latexsym,amsfonts,amssymb,exscale,epsfig}
\usepackage[centertags,sumlimits,intlimits,namelimits,reqno]{amsmath}

\def\theauthor{}
\def\empty{}
\def\theaffiliation{}
\def\preprint#1{
  \thispagestyle{plain}
  \def\theauthor{#1}
  \ifx\theauthor\empty
  \else
    \begin{flushright}{\small #1\par}\end{flushright}
  \fi
  \begin{center}}
\def\title#1{
  {\LARGE #1\par}\vskip 1em}
\def\author#1{
  \ifx\theaffiliation\empty
  \else
    \par
  \fi
  \def\theauthor{#1}\def\theaffiliation{}}
\def\email#1{
  \vskip 1em{\large\theauthor\footnote{\small email: {\tt #1}}\par}\vskip .5em}
\def\affiliation#1{
  \ifx\theaffiliation\empty
    \def\theaffiliation{second}
  \else
  \fi
  {\small\sl #1\par}}
\def\date#1{
  \vskip 1em{(#1)\par}\end{center}\vskip 2em}

\def\acknowledgments{\section*{Acknowledgments}}%
\def\pacs#1{\noindent {\small PACS: #1\par}}%
\def\keywords#1{\noindent {\small keywords: #1\par}}%

\usepackage{amsthm}
\theoremstyle{definition}
\newtheorem{theorem}{Theorem}[section]
\newtheorem{lemma}[theorem]{Lemma}

\newtheorem{definition}[theorem]{Definition}
\newtheorem{remark}[theorem]{Remark}
\newtheorem{example}[theorem]{Example}
\newtheorem{conjecture}[theorem]{Conjecture}

\usepackage{bbm}

\def\R{{\mathbbm R}}
\def\Z{{\mathbbm Z}}

\def\SO{{SO}}

\def\SU{{SU}}

\def\U{{U}}

\def\ie{{\sl i.e.\/}}

\def\etc{{\sl etc.\/}}
\def\cf{{\sl c.f.\/}}
\let\phi=\varphi
\let\theta=\vartheta
\let\epsilon=\varepsilon

\def\id{\mathop{\rm id}\nolimits}

\def\Aut{\mathop{\rm Aut}\nolimits}

\def\ker{\mathop{\rm ker}\nolimits}

\let\hat=\widehat
\let\tilde=\widetilde

\newcommand{\color}[2][c]{}

\numberwithin{equation}{section}

\def\alignidx#1{#1}

\def\nn{\notag}
\def\del{\partial}
\def\emph#1{{\sl #1\/}}
\def\sym#1{{\mathcal #1}}

\usepackage[ps,matrix,arrow,curve]{xy}
\xyoption{dvips} 

\begin{document}
%

\preprint{DAMTP-2003-27}

\title{Higher gauge theory and a non-Abelian generalization\\
       of $2$-form electrodynamics}

\author{Hendryk Pfeiffer}
\email{hpfeiffer@perimeterinstitute.ca}
\affiliation{Perimeter Institute for Theoretical Physics, 35 King Street N, Waterloo, Ontario, N2J 2W9, Canada}
\affiliation{Emmanuel College, St Andrew's Street, Cambridge CB2 3AP, United Kingdom}
\affiliation{DAMTP, Wilberforce Road, Cambridge CB3 0WA, United Kingdom}

\date{April 16, 2003}

%
\begin{abstract}
%

In conventional gauge theory, a charged point particle is described by
a representation of the gauge group. If we propagate the particle
along some path, the parallel transport of the gauge connection acts
on this representation. The Lagrangian density of the gauge field
depends on the curvature of the connection which can be calculated
from the holonomy around (infinitesimal) loops. For Abelian symmetry
groups, say $G=\U(1)$, there exists a generalization, known as
$p$-form electrodynamics, in which $(p-1)$-dimensional charged objects
can be propagated along $p$-surfaces and in which the Lagrangian
depends on a generalized curvature associated with (infinitesimal)
closed $p$-surfaces. In this article, we use Lie $2$-groups and ideas
from higher category theory in order to formulate a discrete gauge
theory which generalizes these models at the level $p=2$ to possibly
non-Abelian symmetry groups. An important feature of our model is that
it involves both parallel transports along paths and generalized
transports along surfaces with a non-trivial interplay of these two
types of variables. Our main result is the geometric picture, namely
the assignment of non-Abelian quantities to geometrical objects in a
coordinate free way. We construct the precise assignment of variables
to the curves and surfaces, the generalized local symmetries and gauge
invariant actions and we clarify which structures can be non-Abelian
and which others are always Abelian. A discrete version of connections
on non-Abelian gerbes is a special case of our construction. Even
though the motivation sketched so far suggests applications mainly in
string theory, the model presented here is also related to spin foam
models of quantum gravity and may in addition provide some insight
into the role of centre monopoles and vortices in lattice QCD.

\end{abstract}


%
\section{Introduction}
%

In the present article, we are concerned with gauge theories in a
discretized formulation on some sort of lattice, triangulation or
cellular decomposition. This includes the standard formulation of
lattice gauge theory, see, for example~\cite{Ro92,MoMu94}, but we can
also think of a smooth manifold with a large collection of embedded
curves and surfaces on which we place the variables of the theory. The
main point is that we keep group elements for the parallel transports
along the curves, but that we do not pass to the differential picture
and do not replace the group by its Lie algebra. The discrete
structure is represented by an abstract simplicial complex, \ie\ a
collection of vertices, edges, triangles, \etc\ which we call
\emph{lattice}. We do not discuss here how these simplices are mapped
to or embedded in some given manifold.

The theory can be classical, in this case we have to define some
action and the variables in the Lagrangian picture, or it can be a
quantum theory in the path integral formulation whose path integral is
the sum or integral over all classical configurations.

Let us consider a gauge theory whose gauge group is any Lie group
$G$. We concentrate on pure gauge fields. There are no dynamical
matter fields. The fundamental variables are taken to be the parallel
transports of the gauge connection along the edges (links) of the
lattice, \ie\ one associates a group element $g_e\in G$ with each edge
$e$. We call the source and target vertices of the edge $s(e)$ and
$t(e)$, respectively.

The gauge connection encodes what happens to charged point particles
if we propagate them on the lattice. A charged particle is a vector
$w\in W$ in some representation $\rho$ of $G$. If we propagate such a
particle from $s(e)$ along the edge $e$ to $t(e)$, then the parallel
transport $g_e$ acts on the vector, $w\mapsto \rho(g_e)w$. Observe
that composition and orientation reversal of the edges correspond
precisely to the group structure of $G$. Since the choice of internal
reference frame, essentially the choice of basis of $W$, is arbitrary,
all physically meaningful quantities are required to be invariant
under \emph{local gauge transformations},
\begin{equation}
\label{eq_localgauge}
  \alignidx{g_e \mapsto h_{s(e)}^{-1}\cdot g_e\cdot h_{t(e)}},
\end{equation}
for each edge $e$. The \emph{generating function} $h$ assigns a group
element $h_v\in G$ to each vertex $v$ and thus parameterizes the local
changes of basis. The action of the theory is a physical quantity and
therefore gauge invariant. It turns out that the easiest way to obtain
a gauge invariant expression is to calculate the product of group
elements around a closed loop and then to evaluate a character of
$G$. The action originally proposed by Wilson makes use of the loop
around an elementary square (plaquette) and calculates the real part
of the character of the fundamental representation of $G$.

Now let us try to generalize the setting. The charged particles are
replaced by line-like objects which are to be propagated along
surfaces. We therefore need additional variables associated with the
faces (plaquettes) of the lattice. As the faces are two-dimensional
objects, we have to deal with various ways of composing individual
faces to larger surfaces, for example putting them below each other
(\emph{vertical composition}) or next to each other (\emph{horizontal
composition}). The algebraic structure we use has to reflect these
geometrical conditions.

It is known that one possible solution is to label the faces with
elements of an Abelian group $G$ and to consider these as the
fundamental variables. The edges are not labeled. For $G=\U(1)$, such
a model is called \emph{$2$-form electrodynamics}, or more generally
$p$-form electrodynamics if the fundamental variables are associated
with the $p$-cells of the lattice (vertices, edges, faces,
$\ldots$). The continuum formulation of the theory for $p=2$ involves
so-called Kalb--Ramond fields~\cite{KaRa74}. The higher level models
were originally introduced in the language of lattice models in
statistical mechanics~\cite{Sa77,Sa80} where they correspond to the
$xy$-model ($p=0$), $\U(1)$-lattice gauge theory ($p=1$), a theory for
an antisymmetric rank-$2$ tensor field ($p=2$) and so on. Their
continuum counterparts have been studied in~\cite{HeTe86}.

Consider the case $p=2$. Can we do any better than just using an
Abelian symmetry group? As long as we insist on colouring only faces,
this is not possible as a classic argument from algebraic topology
shows. But a generalization is possible if we colour both edges and
faces with suitable algebraic structures. The main result of the
present article is the construction of such a generalized $2$-form
gauge theory with its local gauge transformations and gauge invariant
expressions.

We emphasize the geometrical properties of our model, namely that we
have an assignment of non-Abelian quantities to geometrical objects in
a coordinate free way. There exists a considerable literature on the
interplay of Lie algebra valued $1$- and $2$-forms and their extended
`gauge' symmetries, but there is usually~\cite{Te86} no geometrical
interpretation for the non-Abelian $2$-form comparable to the parallel
transport along curves which can be constructed from a connection
$1$-form. In this article, we pursue a complementary approach. We
require a consistent geometrical picture and then deduce how much
freedom we have in choosing the structure of our model.

At the technical level, the only thing we do is to combine two recent
ideas. The first~\cite{Ba02} is the construction of \emph{Lie
$2$-groups} which serve as the algebraic structure in order to label
edges and faces in a consistent way with non-Abelian quantities. The
second~\cite{GrSc01} is the use of ideas from category theory in order
to rephrase lattice gauge theory in a way that does admit the desired
generalization.

A related construction was presented in~\cite{At02} where the discrete
framework is used in order to derive the corresponding continuum
expressions. One can say that the construction presented here is
somewhat more general than a discrete version of a theory of the $1$-
and $2$-connections of non-Abelian gerbes. For mathematical background
on gerbes, see~\cite{BrMe01}. Although we do not use the most general
setting and restrict ourselves to strict Lie $2$-groups, we still have
plenty of examples and our construction includes a discrete version of
non-Abelian gerbes as a special case.

The motivation sketched so far, namely to replace a theory for charged
point particles by a theory for both charged point and charged
line-like particles, seems to be mainly related to string theory. In
fact, the simplest case, Abelian $2$-form electrodynamics or
Kalb--Ramond fields~\cite{KaRa74} play an important role in string
theory so that it is interesting to understand possible
generalizations.

In the context of quantum field theory, one might naively claim that
in four dimensions any theory with a local non-Abelian symmetry is a
gauge theory. Of course, this is not the full truth, and it is of
general interest to understand in conceptual terms which other
theories can have local symmetries.

Beyond these general ideas, we sketch how the model constructed here
might provide further insight into the refined state sum models of
quantum gravity and into the role of centre monopoles and vortices in
lattice QCD. We also comment on the hierarchy of models that
generalize $p$-form electrodynamics to non-Abelian symmetries.

We try to keep the present article self-contained and to make it
readable by physicists who are not yet familiar with category theory.
Therefore we carefully review the required background material. The
article is organized as follows. In Section~\ref{sect_ordinary}, we
rephrase conventional lattice gauge theory in the language of category
theory in order to prepare for the generalization. In
Section~\ref{sect_background}, we recall the argument which forces the
symmetry group to be Abelian if we colour only the faces, and we
present the relevant algebraic structures in order to circumvent
it. The construction of our non-Abelian $2$-form lattice gauge theory
is presented in
Section~\ref{sect_construction}. Section~\ref{sect_examples} contains
some examples and comments on their physical relevance. Finally, in
Section~\ref{sect_conclude}, we discuss open questions and the
relationship with other approaches.

%
\section{Conventional lattice gauge theory}
%
\label{sect_ordinary}

In this section, we review the basic structures of lattice gauge
theory for pure gauge fields and rephrase everything in the language
of category theory. This might seem to be much too complicated at
first sight, but it turns out that the category theoretic language is
the key to the generalization to higher level, including dynamical
variables both at the edges and at the faces of the lattice.

The idea to rephrase lattice gauge theory in the language of category
theory was, to our knowledge, first proposed by Baez~\cite{Ba96} and
then generalized by Grosse and Schlesinger~\cite{GrSc01}. Here we
review this construction in a language adapted to the examples we are
going to present.  For a mathematical introduction to category theory,
see, for example~\cite{Ma98}.

Informally speaking, a category is a collection of points
(\emph{objects}) and arrows between these points (\emph{morphisms})
with enough structure so that we can compose arrows and that we have
\emph{identities}, \ie\ arrows that behave like neutral elements under
composition. The precise definition is as follows. We restrict
ourselves to \emph{small} categories, \ie\ the collections of objects
and morphisms form proper sets.

\begin{definition}
\label{def_category}
A \emph{small category} $\sym{C}=(\sym{C}_0,\sym{C}_1,s,t,\id,\circ)$
consists of a set $\sym{C}_0$ of \emph{objects}, a set $\sym{C}_1$ of
\emph{morphisms} and maps $s\colon\sym{C}_1\to\sym{C}_0$
(\emph{source}), $t\colon\sym{C}_1\to\sym{C}_0$ (\emph{target}),
$\id\colon\sym{C}_0\to\sym{C}_1$ (\emph{identity}) and
$\circ\colon\sym{C}_1\times_{\sym{C}_0}\sym{C}_1\to\sym{C}_1$
(\emph{composition}) such that the following axioms hold,
\begin{gather}
\label{eq_catid}
  s(\id_g)=g=t(\id_g),\\
\label{eq_catsource}
  s(g\circ g^\prime)=s(g),\quad t(g\circ g^\prime)=t(g^\prime),\\
\label{eq_catunit}
  \id_{s(f)}\circ f=f=f\circ\id_{t(f)},\\
\label{eq_catassoc}
  (f_1\circ f_2)\circ f_3=f_1\circ(f_2\circ f_3),
\end{gather}
for all objects $g,g^\prime\in\sym{C}_0$ and morphisms $f\in\sym{C}_1$
and $f_1,f_2,f_3\in\sym{C}_1$ where composable. We have denoted by
\begin{equation}
  \sym{C}_1\times_{\sym{C}_0}\sym{C}_1:=\{\,(f_1,f_2)\in\sym{C}_1\times\sym{C}_1\colon\quad
    s(f_2)=t(f_1)\,\}
\end{equation}
the set of all pairs of composable morphisms. We write $f\colon g_1\to
g_2$ for a morphism $f\in\sym{C}_1$ from the source $g_1=s(f)$ to the
target $g_2=t(f)$. Notice that we read the composition ($\circ$) from
left to right.

A morphism $f\colon g_1\to g_2$ is called \emph{isomorphism} if it has
a two-sided inverse, \ie\ if there exists a morphism $f^{-1}\colon
g_2\to g_1$ such that
\begin{equation}
  f\circ f^{-1}=\id_{g_1},\qquad f^{-1}\circ f=\id_{g_2}.
\end{equation}
A category in which every morphism is an isomorphism, is called
\emph{groupoid}.
\end{definition}

It is instructive to visualize all this using diagrams. For a morphism
$f\colon g_1\to g_2$ we draw an arrow,
\begin{equation}
\begin{aligned}
\xymatrix{
  g_1\ar@/^2ex/[rr]^f&&g_2
}
\end{aligned}
\end{equation}
Composition of morphisms and the identity morphisms are shown in the
following diagram,
\begin{equation}
\begin{aligned}
\xymatrix{
  g_1\ar@(ul,dl)_{\id_{g_1}}\ar[rr]^{f_1}\ar[ddrr]_{f_1\circ f_2}&&g_2\ar[dd]^{f_2}\\
  \\
  &&g_3
}
\end{aligned}
\end{equation}

We will make use of categories for two purposes. First, the gauge
group gives rise to a category and second, the lattice forms a
category as well. This is stated in the following two examples.

\begin{example}
\label{ex_group}
Let $G$ be a Lie group. Then there is a groupoid $\sym{G}^G$
associated with $G$ which has only one object, $\sym{G}^G_0=\{\ast\}$,
and whose morphisms are the group elements,
$\sym{G}^G_1=G$. Composition is the multiplication in $G$. Obviously,
the source and target maps $s,t\colon G\to\{\ast\}$ are trivial, and
the identity map associated with the object $\ast$ is the unit in $G$,
$\id_\ast=1\in G$.
\end{example}

\begin{example}
\label{ex_graph}
Let $(V,E)$ be a directed graph, given by a set $V$ of \emph{vertices}
and a set $E$ of \emph{edges} together with maps $s\colon E\to V$ and
$t\colon E\to V$ indicating the source and target of each edge. Then
there is a category $\sym{C}^{V,E}$ whose objects $\sym{C}^{V,E}_0=V$
are the vertices. The set of morphisms $\sym{C}^{V,E}_1$ comprises
\begin{itemize}
\item
  all edges $e\in E$, 
\item
  for each edge $e\in E$ its orientation reversed counterpart $e^\ast\in E$
  (such that $s(e^\ast)=t(e)$ and $t(e^\ast)=s(e)$),
\item
  for each vertex $v\in V$ one morphism $\id_v$ such that
  $s(\id_v)=v=t(\id_v)$ and
\item
  all formal compositions of these edges, subject to the relations
\begin{gather}
  e\circ e^\ast = \id_{s(e)},\\
  e^\ast\circ e = \id_{t(e)},\\
  \id_{s(e)}\circ e = e = e\circ\id_{t(e)},
\end{gather}
  for all $e\in E$.
\end{itemize}
The maps $s$ and $t$ are given by the directed graph and extended to
compositions such that~\eqref{eq_catsource} holds.
\end{example}

This construction looks complicated at first sight, but the only thing
we have done is to take vertices as objects and edges as morphisms and
make everything else fit the picture.

So far we have introduced two categories. One of them,
$\sym{C}^{V,E}$, encodes the information about the lattice while the
other one, $\sym{G}^G$, is the gauge group. We will see that
configurations of lattice gauge theory are maps from the former to the
latter. Structure preserving maps between categories are known as
\emph{functors} and are defined as follows.

\begin{definition}
Let $\sym{C}=(\sym{C}_0,\sym{C}_1,s,t,\id,\circ)$ and
$\sym{C}^\prime=(\sym{C}^\prime_0,\sym{C}^\prime_1,s^\prime,t^\prime,\id^\prime,\circ^\prime)$
be small categories. A \emph{functor}
$F\colon\sym{C}\to\sym{C}^\prime$ is a pair $(F_0,F_1)$ of maps,
$F_0\colon\sym{C}_0\to\sym{C}^\prime_0$ and
$F_1\colon\sym{C}_1\to\sym{C}^\prime_1$, sending objects to objects
and morphisms to morphisms such that
\begin{gather}
  F_0s(f)=s^\prime(F_1f),\quad F_0t(f)=t^\prime(F_1f),\\
  F_1\id_g=\id^\prime_{F_0g},\\
  F_1(f_1\circ f_2)=F_1f_1\circ^\prime F_1f_2,
\end{gather}
for all objects $g\in\sym{C}_0$ and morphisms $f\in\sym{C}_1$ and
$(f_1,f_2)\in\sym{C}_1\times_{\sym{C}_0}\sym{C}_1$. We are going to
omit the indices $0,1$ of $F_0$, $F_1$ if it is obvious from the
context which map we refer to.
\end{definition}

\begin{example}
\label{ex_config}
Let $(V,E)$ be a directed graph and $G$ be a Lie group. A functor
$F\colon\sym{C}^{V,E}\to\sym{G}^G$ is a pair of maps, $F_0\colon
V\to\{\ast\}$ and $F_1\colon E\to G$. The edges are therefore labeled
by group elements whereas the vertices are not labeled at all so that
a functor $\sym{C}^{V,E}\to\sym{G}^G$ is just a configuration of
lattice gauge theory.

Observe that the identity edges $\id_v$ at each vertex $v\in V$
(see Example~\ref{ex_graph}) are labeled by the group unit $1\in G$. This
implies that orientation reversed edges are assigned the inverse group
element, $F_1(e^\ast)={(F_1e)}^{-1}$. 
\end{example}

So far we have said in the new category theoretic language what a
configuration of lattice gauge theory is, namely a functor
$F\colon\sym{C}^{V,E}\to\sym{G}^G$. From category theory, we know how
to compare two functors, and this concept naturally leads to the
familiar local gauge symmetry.

\begin{definition}
Let $\sym{C}$, $\sym{C}^\prime$ be small categories and $F,\tilde
F\colon\sym{C}\to\sym{C}^\prime$ be functors. A \emph{natural transformation}
$\eta\colon F\Rightarrow \tilde F$ is a map
$\eta\colon\sym{C}_0\to\sym{C}^\prime_1$ that associates with each object
$g\in\sym{C}_0$ a morphism $\eta_g\colon Fg\to \tilde Fg$ in $\sym{C}^\prime$
such that
\begin{equation}
  Ff\circ \eta_{g_2} = \eta_{g_1}\circ\tilde Ff
\end{equation}
holds for all morphisms $f\colon g_1\to g_2$ in $\sym{C}$. This means
that the following diagram commutes,
\begin{equation}
\begin{aligned}
\xymatrix{
  Fg_1\ar[rr]^{Ff}\ar[dd]_{\eta_{g_1}}&&Fg_2\ar[dd]^{\eta_{g_2}}\\
  \\
  \tilde Fg_1\ar[rr]_{\tilde Ff}&&\tilde Fg_2
}
\end{aligned}
\end{equation}
A natural transformation is called \emph{natural equivalence} if
$\eta_g$ is an isomorphism for any $g\in\sym{C}_0$.
\end{definition}

Natural equivalences can now be used in order to compare two
configurations of lattice gauge theory. Let us specialize the
preceding definition to our situation.

\begin{example}
\label{ex_localgauge}
Let $F,\tilde F\colon\sym{C}^{V,E}\to\sym{G}^G$ be two functors. A
natural equivalence $\eta\colon F\Rightarrow \tilde F$ is a map
$\eta\colon V\to G$ such that for each edge $e\colon v\to w$, the
following diagram commutes,
\begin{equation}
\label{eq_natlgt}
\begin{aligned}
\xymatrix{
  \ast\ar[rr]^{Fe}\ar[dd]_{\eta_v}&&\ast\ar[dd]^{\eta_w}\\
  \\
  \ast\ar[rr]_{\tilde Fe}&&\ast
}
\end{aligned}
\end{equation}
In the category $\sym{G}^G$ in which composition is the group product,
this means that
\begin{equation}
  \tilde Fe=\eta_v^{-1}\cdot Fe\cdot\eta_w,
\end{equation}
so that the configuration of lattice gauge theory given by $\tilde F$
is locally gauge equivalent to the configuration given by $F$,
\cf~\eqref{eq_localgauge}. The map $V\to G, v\mapsto \eta_v$ in the
definition of the natural equivalence plays the role of the generating
function.
\end{example}

\begin{figure}[t]
\begin{center}
\input{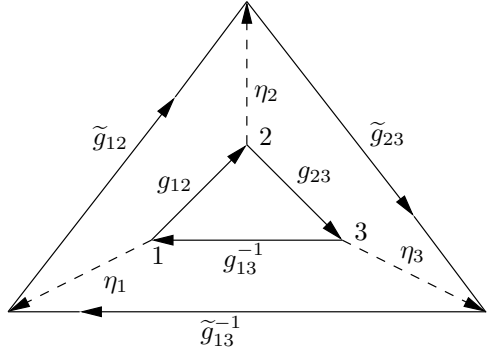}
\end{center}
\caption{\label{fig_triangle}%
The holonomy $\alignidx{g_{12}g_{23}g_{13}^{-1}}$ around some triangle
$(1,2,3)$. The inner triangle is labeled by a functor
$F\colon\sym{C}^{V,E}\to\sym{G}^G$, the outer triangle by $\tilde
F$. The functors $F$ and $\tilde F$ are related by a natural
equivalence $\eta$.  }
\end{figure}

Which expressions are gauge invariant and can therefore correspond to
physically meaningful quantities? We know that the simplest such
expressions are Wilson loops, group characters evaluated at the
holonomy around loops.

Consider Figure~\ref{fig_triangle}. It shows a triangle $(1,2,3)$
labeled in two ways. The inner triangle is labeled by a functor
$F\colon\sym{C}^{V,E}\to\sym{G}^G$, the outer one by $\tilde F$. We
have set $g_{ij}:=Fe_{ij}$ where $e_{ij}$ are the edges such that
$s(e_{ij})=i$ and $t(e_{ij})=j$ and similarly $\tilde g_{ij}:=\tilde
Fe_{ij}$. The figure contains three commutative squares of the
form~\eqref{eq_natlgt}. Commutativity implies that
\begin{equation}
  \alignidx{{\tilde g}_{12}{\tilde g}_{23}{\tilde g}_{13}^{-1} =
    \eta_1^{-1}g_{12}g_{23}g_{13}^{-1}\eta_1},
\end{equation}
so that any group character of the holonomy,
$\chi(\alignidx{g_{12}g_{23}g_{13}^{-1}})$, is a gauge invariant
quantity.

Of course, in standard lattice gauge theory, the properties mentioned
so far are much easier to verify in direct computations. The category
theoretic language presented here, however, provides a structural
framework which would have allowed us to derive all these properties
in a systematic fashion even if we had not known them in advance. We
will exploit this conceptual advantage when we generalize lattice
gauge theory to the next level, colouring both edges and faces with
dynamical variables. In Section~\ref{sect_construction}, we will
encounter the higher level analogues of all the diagrams used here.
Without any help from category theory it would hardly be possible to
guess the appropriate assignment of variables and the relevant
symmetries.

Our plan for the following section is to review suitable generalizations
for the notions of category, functor and natural transformation in
order to pass on to the next level, including interesting examples for
which we can perform explicit computations. The higher level
generalizations are known as $2$-categories, $2$-functors, \etc.

%
\section{Mathematical background}
%
\label{sect_background}

\subsection{The Eckmann--Hilton argument}
\label{sect_eckmann}

\begin{figure}[t]
\begin{center}
\input{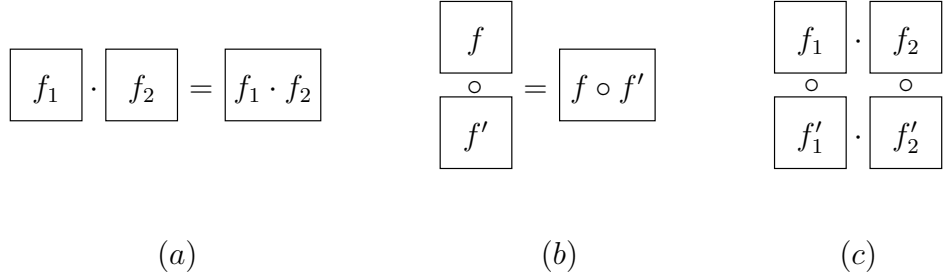}
\end{center}
\caption{\label{fig_eckmann}%
  (a) Horizontal composition of faces is denoted by a dot
  ($\cdot$). (b)~Vertical composition is indicated by a little circle
  ($\circ$) which is read from left to right in our equations. (c)
  Parentheses are not necessary provided the exchange
  law~\eqref{eq_exchange} holds.}
\end{figure}

Let us first review the argument of Eckmann and Hilton~\cite{EcHi62}
which explains why we are forced to use Abelian groups as long as we
colour only the faces.

We assume that the plaquettes of the lattice are labeled by the
elements $f\in G$ of some algebraic structure $G$. We further assume
that there are two composition laws as illustrated in
Figure~\ref{fig_eckmann}. \emph{Horizontal composition} is denoted by
a dot ($\cdot$), \emph{vertical composition} by a circle
($\circ$). Finally, we assume that both compositions have two-sided
units, denoted by $1_\cdot$ and $1_\circ$. On larger lattices, there
will occur mixed compositions such as that shown in
Figure~\ref{fig_eckmann}(c). We require that this composition is well
defined without parentheses, \ie\ it does not depend on the ordering
by which horizontal and vertical compositions are performed. This
means that the \emph{exchange law},
\begin{equation}
\label{eq_exchange}
  (f_1\cdot f_2)\circ (f_1^\prime\cdot f_2^\prime)
  = (f_1\circ f_1^\prime)\cdot(f_2\circ f_2^\prime),
\end{equation}
is satisfied.

This implies first that the units actually agree because
\begin{equation}
  1_\circ = 1_\circ\circ 1_\circ=(1_\circ\cdot 1_\cdot)\circ(1_\cdot\cdot 1_\circ)
  = (1_\circ\circ 1_\cdot)\cdot(1_\cdot\circ 1_\circ)
  = 1_\cdot\cdot 1_\cdot=1_\cdot
\end{equation}
Therefore we can write $1=1_\cdot=1_\circ$. The exchange law further
implies that both compositions agree because for any $f,g\in G$,
\begin{equation}
  f\cdot g = (f\circ 1)\cdot (1\circ g)=(f\cdot 1)\circ (1\cdot g)
  =f\circ g,
\end{equation}
and finally that this composition is Abelian,
\begin{equation}
\label{eq_abelian}
  f\cdot g = (1\circ f)\cdot (g\circ 1)=(1\cdot g)\circ (f\cdot 1)
  = g\circ f = g\cdot f.
\end{equation}
Drawing the diagrams corresponding to~\eqref{eq_abelian}, it becomes
obvious that the two-dimensionality of the situation allows $f$ and
$g$ to move around in the plane and thereby to change places.

If we wish to escape this Abelianness, we have to change some of the
initial assumptions. There are \emph{a priori} various conceivable
approaches. Since we aim for a setting similar to lattice gauge
theory, we have to require that Figure~\ref{fig_eckmann}(c) is
well-defined without parentheses so that the exchange
law~\eqref{eq_exchange} is not in question. It turns out that it is a
viable strategy to colour both edges and faces with different
algebraic structures in such a way that there is a non-trivial
interplay and that the identities no longer agree.

An observation related to the Eckmann--Hilton argument was made
independently by Teitelboim~\cite{Te86} in a physical context. The
idea is as follows. In conventional gauge theory, there is the notion
of a `path ordered product' by which one defines the parallel
transport along some curve and which is independent of the
parameterization and thus a geometrical quantity. At higher level,
however, one is forced to use Abelian labels because it seems to be
impossible to define a `surface ordered product' for generic
non-Abelian quantities in a way that is independent of the choice of
coordinates.

\subsection{Lie $2$-groups}

In order to sidestep the Eckmann--Hilton argument, we follow ideas
from higher category theory as explained in~\cite{Ba02}. The picture
is as follows. The edges, going from one vertex to another, are
labeled with elements $g_1,g_2$ from one algebraic
structure. Composition of edges has to be reflected in this algebraic
structure,
\begin{equation}
\label{eq_compose}
\begin{aligned}
\xymatrix{
  \bullet\ar@/^1ex/[r]^{g_1}&\bullet\ar@/^1ex/[r]^{g_2}&\bullet&\longmapsto&\bullet\ar@/^1ex/[rr]^{g_1\cdot g_2}&&\bullet.
}
\end{aligned}
\end{equation}
We therefore require an associative product $g_1\cdot g_2$. In
addition to this, there are faces, going from one edge to another,
which are labeled with elements $f$ from another algebraic
structure. Here we use \emph{bi-gons} as the fundamental faces, \ie\
their boundary consists of only two edges that have the same source
and target vertex,
\begin{equation}
\label{eq_bigon}
\begin{aligned}
\xymatrix{
  \bullet\ar@/^2ex/[rr]^{g}="g1"\ar@/_2ex/[rr]_{g^\prime}="g2"&&\bullet.
  \ar@{=>}_f "g1"+<0ex,-2ex>;"g2"+<0ex,2ex>
}
\end{aligned}
\end{equation}
These bi-gons can then be composed horizontally,
\begin{equation}
\label{eq_horizontal}
\begin{aligned}
\xymatrix{
  \bullet\ar@/^2ex/[rr]^{g_1}="g1"\ar@/_2ex/[rr]_{g_1^\prime}="g2"&&
  \bullet\ar@/^2ex/[rr]^{g_2}="g3"\ar@/_2ex/[rr]_{g_2^\prime}="g4"&&\bullet
  \ar@{=>}_{f_1} "g1"+<0ex,-2ex>;"g2"+<0ex,2ex>
  \ar@{=>}_{f_2} "g3"+<0ex,-2ex>;"g4"+<0ex,2ex>
}
\qquad\longmapsto\qquad
\xymatrix{
  \bullet\ar@/^2ex/[rrr]^{g_1\cdot g_2}="g5"\ar@/_2ex/[rrr]_{g_1^\prime\cdot g_2^\prime}="g6"&&&\bullet,
  \ar@{=>}_{f_1\cdot f_2} "g5"+<0ex,-2ex>;"g6"+<0ex,2ex>
}
\end{aligned}
\end{equation}
and vertically,
\begin{equation}
\label{eq_vertical}
\begin{aligned}
\xymatrix{
  \bullet\ar@/^4ex/[rrr]^{g}="g1"\ar[rrr]^(0.35){g^\prime}\ar@{}[rrr]|{}="g2"
   \ar@/_4ex/[rrr]_{g^{\prime\prime}}="g3"&&&\bullet
  \ar@{=>}^f "g1"+<0ex,-2ex>;"g2"+<0ex,1ex>
  \ar@{=>}^{f^\prime} "g2"+<0ex,-1ex>;"g3"+<0ex,2ex>
}
\qquad\longmapsto\qquad
\xymatrix{
  \bullet\ar@/^3ex/[rrr]^{g}="g4"\ar@/_3ex/[rrr]_{g^{\prime\prime}}="g5"&&&\bullet.
  \ar@{=>}_{f\circ f^\prime} "g4"+<0ex,-2ex>;"g5"+<0ex,2ex>
}
\end{aligned}
\end{equation}
The mixed composition,
\begin{equation}
\label{eq_mixed}
\begin{aligned}
\xymatrix{
  \bullet\ar@/^4ex/[rrr]^{g_1}="g1"\ar[rrr]^(0.35){g_1^\prime}\ar@{}[rrr]|{}="g2"
   \ar@/_4ex/[rrr]_{g_1^{\prime\prime}}="g3"&&&
  \bullet\ar@/^4ex/[rrr]^{g_2}="g4"\ar[rrr]^(0.35){g_2^\prime}\ar@{}[rrr]|{}="g5"
   \ar@/_4ex/[rrr]_{g_2^{\prime\prime}}="g6"&&&\bullet,
  \ar@{=>}^{f_1} "g1"+<0ex,-2ex>;"g2"+<0ex,1ex>
  \ar@{=>}^{f_1^\prime} "g2"+<0ex,-1ex>;"g3"+<0ex,2ex>
  \ar@{=>}^{f_2} "g4"+<0ex,-2ex>;"g5"+<0ex,1ex>
  \ar@{=>}^{f_2^\prime} "g5"+<0ex,-1ex>;"g6"+<0ex,2ex>
}
\end{aligned}
\end{equation}
is required to be independent of the ordering of the horizontal and
vertical compositions which is precisely stated by the exchange law,
\begin{equation}
\label{eq_exchange2}
  (f_1\circ f_1^\prime)\cdot (f_2\circ f_2^\prime)
  = (f_1\cdot f_2)\circ(f_1^\prime\cdot f_2^\prime).
\end{equation}

Lie $2$-groups as introduced by Baez~\cite{Ba02} form a suitable
structure with these properties. Note that the Eckmann--Hilton
argument can be avoided here because we have labeled the edges by
$g_1,g_2$ and we perform some sort of computation $g_1\cdot g_2$
whenever edges are composed. The units for the vertical composition
($\circ$) will in general depend on the $g_i$ and therefore need not
agree. When we list examples below, we mention in which cases the
Eckmann--Hilton argument applies and in which it does not.

Strict Lie $2$-groups can be obtained by a standard construction in
category theory. A strict Lie $2$-group is an \emph{internal category}
in the category of Lie groups. This means that we take the definition
of a small category (Definition~\ref{def_category}) and systematically
replace the word `set' by `Lie group' and `map' by `Lie group
homomorphism'.

\begin{definition}
\label{def_lietwo}
A \emph{strict Lie $2$-group}
$\sym{C}=(\sym{C}_0,\sym{C}_1,s,t,\id,\circ)$ consists of two Lie
groups $\sym{C}_0$ and $\sym{C}_1$ with Lie group homomorphisms
$s\colon\sym{C}_1\to\sym{C}_0$, $t\colon\sym{C}_1\to\sym{C}_0$,
$\id\colon\sym{C}_0\to\sym{C}_1$ and
$\circ\colon\sym{C}_1\times_{\sym{C}_0}\sym{C}_1\to\sym{C}_1$ such
that the axioms~\eqref{eq_catid}--\eqref{eq_catassoc} hold.
\end{definition}

\begin{remark}
\begin{enumerate}
\item
  For elements $g_1,g_2\in\sym{C}_0$ we write single arrows while
  elements $f\in\sym{C}_1$, $f\colon g_1\Rightarrow g_2$ are
  visualized by double arrows, \cf~\eqref{eq_bigon}.
\item
  In a strict Lie $2$-group, we have two composition laws for the set
  $\sym{C}_1$. One of them is the group product as $\sym{C}_1$ is now
  a Lie group. We call it \emph{horizontal composition}
  (see~\eqref{eq_horizontal}) and write a dot ($\cdot$) which we
  sometimes even omit. Observe that the source of a horizontal
  composition is the product of the sources because
  $s\colon\sym{C}_1\to\sym{C}_0$ is a group homomorphism, \etc. The
  other composition law is denoted by a circle ($\circ$). It
  originates from the definition of a category and is explicitly
  listed in Definition~\ref{def_lietwo}. This is called \emph{vertical
  composition} (see~\eqref{eq_vertical}).
\item
  Notice that the exchange law~\eqref{eq_exchange2} holds because the
  composition map
  $\circ\colon\sym{C}_1\times_{\sym{C}_0}\sym{C}_1\to\sym{C}_1$ is a
  Lie group homomorphism.
\item
  A generalization of strict $2$-groups is is provided by \emph{weak}
  and \emph{coherent $2$-groups}~\cite{La02}. We restrict ourselves to
  the strict case.
\end{enumerate}
\end{remark}

Concerning the structure of strict Lie $2$-groups, we quote the
following results from~\cite{Ba02,Fo02}.

\begin{lemma}
Let $\sym{C}$ be a strict Lie $2$-group. In particular we have Lie
group homomorphisms $s\colon\sym{C}_1\to\sym{C}_0$ and
$\id\colon\sym{C}_0\to\sym{C}_1$ such that $s(\id(g))=g$ for all
$g\in\sym{C}_0$,
\begin{equation}
\begin{aligned}
\xymatrix{
  \sym{C}_1\ar@<1ex>[rr]^{s}&&\sym{C}_0\ar@<1ex>[ll]^{\id}
}
\end{aligned}
\end{equation}

\begin{enumerate}
\item
  Each element $f\in\sym{C}_1$ has a unique decomposition of the form
  $f=h\cdot \id_g$ where $g=s(f)\in\sym{C}_0$ and $h\in\ker
  s\unlhd\sym{C}_1$\footnote{The notation $A\unlhd B$ indicates that
  $A$ is a normal subgroup of $B$.}.
\item
  There is an isomorphism of Lie groups $\ker
  s\rtimes\sym{C}_0\cong\sym{C}_1$, given by $(h,g)\mapsto
  h\cdot\id_g$. The semi-direct product $\ker s\rtimes\sym{C}_0$ is
  defined so that
\begin{equation}
\label{eq_semimult}
  (h_1,g_1)\cdot(h_2,g_2):=(h_1\alpha(g_1)[h_2],g_1g_2),
\end{equation}
  for $h_1,h_2\in\ker s$ and $g_1,g_2\in\sym{C}_0$, where we set
  $\alpha(g)[h]:=\id_g h\id_{g^{-1}}$.
\item
  The vertical composition of elements of $\sym{C}_1$ is already fixed
  by the structure described so far. In the notation of the
  semi-direct product, it reads
\begin{equation}
  (h_1,g_1)\circ(h_2,g_2) = (h_2h_1,g_1),
\end{equation}
  for $h_1,h_2\in\ker t$ and $g_1,g_2\in\sym{C}_0$, whenever
  composable.
\end{enumerate}
\end{lemma}

\subsection{Lie crossed modules}

For an introduction to internal categories in the category of groups,
see~\cite{Fo02}. One can prove~\cite{Ba02,Fo02} (also see
Chapter~XII.8 of~\cite{Ma98}) that strict Lie $2$-groups are in
one-to-one correspondence (up to isomorphism) with \emph{Lie crossed
modules}. The language of Lie crossed modules is most convenient in
order to construct interesting examples of Lie $2$-groups. They are
defined as follows.

\begin{definition}
A \emph{Lie crossed module} $(G,H,t,\alpha)$ consists of two Lie
groups $G$ and $H$ with Lie group homomorphisms $t\colon H\to G$ and
$\alpha\colon G\to\Aut(H)$ (\ie\ $\alpha$ is an action of $G$ on $H$
that is compatible with the group structure of $H$; we write it as
$\alpha(g)[h]$ for $g\in G$, $h\in H$), such that
\begin{gather}
t(\alpha(g)[h])=g t(h) g^{-1},\\
\alpha(t(h))[h^\prime]=h h^\prime h^{-1},
\end{gather}
for all $g\in G$ and $h,h^\prime\in H$.
\end{definition}

\begin{theorem}[see \cite{Ba02,Fo02}]
\label{thm_liecrossed}
\begin{enumerate}
\item
  Let $\sym{C}$ be a strict Lie $2$-group. Then there is a Lie crossed
  module $(G,H,t,\alpha)$ defined as follows. Define $G:=\sym{C}_0$
  and $H:=\ker s$ to be the kernel of the source homomorphism. The map
  $t\colon H\to G$ is defined to be the restriction $t|_H$ of the
  target homomorphism. Finally, $\alpha(g)[h]:=\id_g h\id_{g^{-1}}$.
\item
  Let $(G,H,t,\alpha)$ be a Lie crossed module. Then there is a strict
  Lie $2$-group $\sym{C}$ defined as follows. Set $\sym{C}_0:=G$ and
  $\sym{C}_1:=H\rtimes G$, the semi-direct product given
  by~\eqref{eq_semimult} using the map $\alpha$ provided by the Lie
  crossed module. The source, target, identity and composition maps
  are defined as follows,
\begin{alignat}{2}
  s&\colon H\rtimes G\to G,&\quad& (h,g)\mapsto g,\\
  t&\colon H\rtimes G\to G,&\quad& (h,g)\mapsto t(h)g,\\
  \id&\colon G\to H\rtimes G,&\quad& g\mapsto (1,g),
\end{alignat}
\begin{equation}
  (h,g)\circ(h^\prime,g^\prime):=(h^\prime h,g),
\end{equation}
  for $g,g^\prime\in G$ and $h,h^\prime\in H$, whenever composable.
\end{enumerate}
\end{theorem}

\begin{remark}
Let us now assume that we have constructed a strict Lie $2$-group from
a Lie crossed module $(G,H,t,\alpha)$ as in the preceding theorem.
\begin{enumerate}
\item
  Horizontal composition is given by the product in $H\rtimes G$. In
  particular, horizontal composition has the identity $(1,1)$ and
  inverses
\begin{equation}
\label{eq_semiinverse}
  {(h,g)}^{-1}=(\alpha(g^{-1})[h^{-1}],g^{-1}).
\end{equation}
\item
  Any pair of elements of $H\rtimes G$ can be composed horizontally.
\item
  Elements $(h_2,g_2),(h_1,g_1)\in H\rtimes G$ are vertically
  composable to $(h_1,g_1)\circ(h_2,g_2)$ if and only if
  $g_2=t(h_1)g_1$.
\item
  Vertical composition has the units $\id_g=(1,g)$, $g\in G$, and is
  always invertible,
\begin{equation}
  (h,g)^\ast = (h^{-1},t(h)g),
\end{equation}
  where we write a star ($\ast$) for the inverse with respect to
  $\circ$.
\item
  The group $H\unlhd H\rtimes G$ parameterizes all morphisms that have
  the unit $1\in G$ as their source. All morphisms whose source and
  target agree, are parameterized by $\ker t\unlhd H$.
\end{enumerate}
\end{remark}

For proofs and more technical details, see~\cite{Ba02,Fo02,Ma98}. A
list of examples can be found in~\cite{Ba02} some of which we mention
here.

\begin{example}
\label{ex_examples}
\begin{enumerate}
\item
  The \emph{trivial $2$-group}. $G$ is any Lie group and $H$ is
  trivial. This example is uninteresting except for the fact that it
  confirms that ordinary Lie groups form a special case of strict Lie
  $2$-groups.
\item
  The \emph{purely Abelian $2$-group}. $G$ is trivial. In this case
  $H$ is Abelian by the Eckmann--Hilton argument. This example gives
  rise to Abelian $2$-form electrodynamics.
\item
  $H$ is an Abelian Lie group on which $G$ acts by some action
  $\alpha\colon G\to\Aut H$. The map $t$ is
  trivial\footnote{Triviality of $t$ already implies that $H$ is
  Abelian by the Eckmann--Hilton argument. This example nevertheless
  provides a non-trivial generalization of $2$-form electrodynamics
  because it involves an interplay of $G$ and $H$ via the action
  $\alpha$.}, $t\colon H\to G,h\mapsto 1$.
\item
  The \emph{Euclidean $2$-groups}. As a special case of (3.) we can
  choose $H=V$ to be some $\R$-vector space (translations) equipped
  with a non-degenerate symmetric bilinear form $\eta$. The group
  $G=\SO(V,\eta)$ (rotations) acts on $V$. From this we obtain the
  \emph{Poincar\'e $2$-group}~\cite{Ba02} which is employed in the
  refined state sum model of~\cite{CrYe03a} if we choose $V=\R^{3+1}$
  with the scalar product of Minkowski space.
\item
  The \emph{automorphism $2$-group}. This is finally an example with
  non-trivial $t$. Choose any Lie group $H$ and $G:=\Aut H$ to be its
  group of automorphisms. The action of $G$ on $H$ is the action by
  the particular automorphism, and $t\colon H\to G$ assigns the inner
  automorphism, conjugation by $h$, to each element $h\in H$. This
  example gives rise to the lattice version of a theory involving the
  connections on non-Abelian gerbes. This Lie $2$-group is related to
  the bi-torsors that are usually employed in the study of non-Abelian
  gerbes, see~\cite{Ba02} for details.
\item
  Many examples of finite and discrete crossed modules are known in
  algebraic topology where crossed modules are a standard tool. For a
  recent survey, see, for example~\cite{Br99}.
\end{enumerate}
\end{example}

Summarizing this section, we can say that the notion of a strict Lie
$2$-group is useful in order make the categorical structure
transparent while the notion of a Lie crossed module provides us with
particular examples and allows us to perform calculations.

\subsection{Suitable $2$-categories}

We have already seen that Lie $2$-groups provide two compositions with
the required identities such that the relevant diagrams
(\cf~\eqref{eq_mixed}) can be drawn and such that the Eckmann--Hilton
argument can be avoided. Before we can generalize the constructions of
Section~\ref{sect_ordinary}, however, we need higher level analogues
of category, functor and natural equivalence. We therefore have to
climb up by one level and introduce the notion of $2$-categories.

A $2$-category $\sym{C}$ is a category `enriched in Cat', \ie\ for
each pair of objects $(x,y)$, we have now a category $\sym{C}(x,y)$
rather than just the set of all morphisms $x\to y$. The objects in
$\sym{C}(x,y)$ are the morphisms of $\sym{C}$ and the morphisms of
$\sym{C}(x,y)$ are new data. They are called \emph{$2$-morphisms}. The
definition of Mac Lane~\cite{Ma98} is of this type. We restrict
ourselves to the \emph{strict} case, \ie\ equalities of morphisms are
satisfied exactly and are not \emph{weakened} to hold only up to
$2$-isomorphism. For more details on $2$-categories, see Chapter XII.3
of~\cite{Ma98} and~\cite{Gr74}. In the following, we remove one level
of abstraction from this definition and write down the conditions in
detail.

\begin{definition}
A \emph{small strict $2$-category} consists of sets $\sym{C}_0$
(\emph{objects}), $\sym{C}_1$ (\emph{morphisms}) and $\sym{C}_2$
(\emph{$2$-morphisms}) together with various maps satisfying axioms as
follows.
\begin{enumerate}
\item
  Maps $s^{(1)}\colon\sym{C}_1\to\sym{C}_0$,
  $t^{(1)}\colon\sym{C}_1\to\sym{C}_0$,
  $\id^{(1)}\colon\sym{C}_0\to\sym{C}_1$ and
  $\cdot\colon\sym{C}_1\times_{\sym{C}_0}\sym{C}_1\to\sym{C}_1$
  such that $(\sym{C}_0,\sym{C}_1,s^{(1)},t^{(1)},\id^{(1)},\cdot)$
  forms a small category (Definition~\ref{def_category}). We have
  denoted by
\begin{equation}
  \sym{C}_1\times_{\sym{C}_0}\sym{C}_1:=\{\,(g_1,g_2)\in\sym{C}_1\times\sym{C}_1
    \colon\quad s^{(1)}(g_2)=t^{(1)}(g_1)\,\}
\end{equation}
  the pairs of horizontally composable morphisms. Compositions in this
  category are visualized by diagrams such as~\eqref{eq_compose}.
\item
  Maps $s^{(2)}\colon\sym{C}_2\to\sym{C}_1$,
  $t^{(2)}\colon\sym{C}_2\to\sym{C}_1$,
  $\id^{(2)}\colon\sym{C}_2\to\sym{C}_1$ and
  $\circ\colon\sym{C}_2\times_{\sym{C}_1}\sym{C}_2\to\sym{C}_2$ such
  that $(\sym{C}_1,\sym{C}_2,s^{(2)},t^{(2)},\id^{(2)},\circ)$ forms
  a small category. We have denoted by
\begin{equation}
  \sym{C}_2\times_{\sym{C}_1}\sym{C}_2:=\{\,(f_1,f_2)\in\sym{C}_1\times\sym{C}_2
    \colon\quad s^{(2)}(f_2)=t^{(2)}(f_1)\,\}
\end{equation}
  the pairs of vertically composable $2$-morphisms.
\item
  Axioms
\begin{equation}
  t^{(1)}(s^{(2)}(f)) = t^{(1)}(t^{(2)}(f)),\qquad
  s^{(1)}(s^{(2)}(f)) = s^{(1)}(t^{(2)}(f)),
\end{equation}
  stating that source $s^{(2)}(f)$ and target $t^{(2)}(f)$ of any
  $2$-morphism $f\in\sym{C}_2$ are parallel morphisms, \ie\ that
  $2$-morphisms are bi-gons as in~\eqref{eq_bigon}. The composition
  ($\circ$) listed under item (2.) is therefore visualized by diagrams
  such as~\eqref{eq_vertical}.
\item
  A map
  $\cdot\colon\sym{C}_2\times_{\sym{C}_0}\sym{C}_2\to\sym{C}_2$
  (horizontal composition of $2$-morphisms), where
\begin{equation}
  \sym{C}_2\times_{\sym{C}_0}\sym{C}_2:=\{\,(f_1,f_2)\in\sym{C}_2\times\sym{C}_2
    \colon\quad t^{(1)}(s^{(2)}(f_1))=s^{(1)}(s^{(2)}(f_2))\,\}
\end{equation}
  denotes the set of all pairs of horizontally composable
  $2$-morphisms. This composition is shown in
  diagram~\eqref{eq_horizontal}.
\item
  Further axioms\footnote{In more abstract terms they state that
  $(\sym{C}_0,\sym{C}_2,s^{(2)}\circ s^{(1)},t^{(2)}\circ
  t^{(1)},\id^{(1)}\circ\id^{(2)},\cdot)$ forms a small category and
  that $s^{(2)}$, $t^{(2)}$ and $\id^{(2)}$ give rise to functors
  between this category and the one formed by $\sym{C}_0$ and
  $\sym{C}_1$.},
\begin{gather}
  s^{(2)}(f_1\cdot f_2) = s^{(2)}(f_1)\cdot s^{(2)}(f_2),\\
  t^{(2)}(f_1\cdot f_2) = t^{(2)}(f_1)\cdot t^{(2)}(f_2),\\
  \id^{(2)}(\id^{(1)}(s^{(1)}(s^{(2)}(f))))\cdot f = f =
    f\cdot \id^{(2)}(\id^{(1)}(t^{(1)}(t^{(2)}(f)))),\\
  (f_1\cdot f_2)\cdot f_3 = f_1\cdot (f_2\cdot f_3),
\end{gather}
  for any $f\in\sym{C}_2$ and horizontally composable
  $f_1,f_2,f_3\in\sym{C}_2$, as well as,
\begin{gather}
  \id^{(2)}(g_1)\cdot \id^{(2)}(g_2)=\id^{(2)}(g_1\cdot g_2),\\
  (f_1\circ f_1^\prime)\cdot (f_2\circ f_2^\prime)
  = (f_1\cdot f_2)\circ(f_1^\prime\cdot f_2^\prime),
\end{gather}
  for any $g_1,g_2\in\sym{C}_1$ and
  $f_1,f_1^\prime,f_2,f_2^\prime\in\sym{C}_2$ whenever they are
  composable.
\end{enumerate}
 A $2$-morphism $f\colon g_1\Rightarrow g_2$ is called
 \emph{$2$-isomorphism} if it has a two-sided inverse, \ie\ if there
 exists a $2$-morphism $f^\ast\colon g_2\Rightarrow g_1$ such that,
\begin{equation}
  f\circ f^\ast=\id^{(2)}_{g_1},\qquad f^\ast\circ f=\id^{(2)}_{g_2}.
\end{equation}
A strict $2$-category in which all morphisms and all $2$-morphisms are
isomorphisms, is called a \emph{strict $2$-groupoid}.
\end{definition}

Recall that a group gives rise to a groupoid with one object
(Example~\ref{ex_group}). In a similar way, a strict Lie $2$-group
provides us with a strict $2$-groupoid with one object. In the
following, we use the language of Lie crossed modules in order to
explicitly describe this strict Lie $2$-group.

\begin{example}
Let $(G,H,t,\alpha)$ be a Lie crossed module with the definitions used
in Theorem~\ref{thm_liecrossed}, item (2.). Then there is a small
strict $2$-groupoid $\sym{G}^{G,H}$ defined as
follows\footnote{Strictly speaking, we should call it
$\sym{G}^{G,H,t,\alpha}$.}. Set $\sym{G}^{G,H}_0:=\{\ast\}$,
$\sym{G}^{G,H}_1:=G$ and $\sym{G}^{G,H}_2:=H\rtimes G$, the
semi-direct product given by~\eqref{eq_semimult}.

The maps $s^{(1)}$ and $t^{(1)}$ are trivial, $\id_\ast^{(1)}=1\in G$
and the composition of morphisms $(\cdot)$ is the multiplication in
$G$. This agrees precisely with Example~\ref{ex_group}.

The other maps are defined as follows,
\begin{alignat}{2}
  s^{(2)}&\colon H\rtimes G\to G,&\quad& (h,g)\mapsto g,\\
  t^{(2)}&\colon H\rtimes G\to G,&\quad& (h,g)\mapsto t(h)g,\\
  \id^{(2)}&\colon G\to H\rtimes G,&\quad& g\mapsto (1,g),
\end{alignat}
and, whenever composable,
\begin{equation}
  (h,g)\circ(h^\prime,g^\prime):=(h^\prime h,g),
\end{equation}
for all $g,g^\prime\in G$ and $h,h^\prime\in H$. The horizontal
composition of $2$-morphisms ($\cdot$) is the product in $H\rtimes G$,
\cf~\eqref{eq_semimult}.
\end{example}

This $2$-groupoid with one object, obtained from a strict Lie
$2$-group, is our generalized notion of gauge group. In order to
specify configurations and local gauge transformations, we have to
introduce structure preserving maps between $2$-categories, called
\emph{$2$-functors}, and suitable natural equivalences.

\begin{definition}
Let $\sym{C}$ and $\sym{C}^\prime$ be small strict $2$-categories. A
\emph{strict $2$-functor} $F\colon\sym{C}\to\sym{C}^\prime$ is a triple of maps
$F_0\colon\sym{C}_0\to\sym{C}^\prime_0$,
$F_1\colon\sym{C}_1\to\sym{C}^\prime_1$, and
$F_2\colon\sym{C}_2\to\sym{C}^\prime_2$, sending objects to objects,
morphisms to morphisms and $2$-morphisms to $2$-morphisms, such that
the following conditions hold,
\begin{enumerate}
\item
  $(F_0,F_1)$ is a functor
  $(\sym{C}_0,\sym{C}_1,s^{(1)},t^{(1)},\id^{(1)},\cdot)
  \to(\sym{C}^\prime_0,\sym{C}^\prime_1,{s^\prime}^{(1)},{t^\prime}^{(1)},{\id^\prime}^{(1)},\cdot^\prime)$,
\item
  $(F_1,F_2)$ is a functor
  $(\sym{C}_1,\sym{C}_2,s^{(2)},t^{(2)},\id^{(2)},\circ)
  \to(\sym{C}^\prime_1,\sym{C}^\prime_2,{s^\prime}^{(2)},{t^\prime}^{(2)},{\id^\prime}^{(2)},\circ^\prime)$,
\item
  $F_2(f_1\cdot f_2)=F_2f_1 \cdot^\prime F_2f_2$ for any
  $(f_1,f_2)\in\sym{C}_2\times_{\sym{C}_0}\sym{C}_2$.
\end{enumerate}
\end{definition}

Given any two parallel $2$-functors between $2$-categories, we seek
the notion of a natural transformation in order to compare these
$2$-functors. There are various flavours of these defined in the
literature. We need what is usually called \emph{pseudo-natural
transformation}. It is a quasi-natural transformation in the
terminology of~\cite{Gr74} in which all $2$-morphisms are isomorphisms.

\begin{definition}
\label{def_pseudonat}
Let $\sym{C}$, $\sym{C}^\prime$ be small strict $2$-categories and $F,\tilde
F\colon\sym{C}\to\sym{C}^\prime$ be parallel strict $2$-functors. A
\emph{pseudo-natural transformation} $\eta\colon F\Rightarrow\tilde F$
is a pair of maps $\eta\colon\sym{C}_0\to\sym{C}^\prime_1$ and
$\eta\colon\sym{C}_1\to\sym{C}^\prime_2$ associating a morphism $\eta_x\colon
Fx\to\tilde Fx$ to each object $x\in\sym{C}_0$ and a $2$-isomorphism
$\eta_g\colon Fg\cdot\eta_{t(g)}\Rightarrow \eta_{s(g)}\cdot\tilde Fg$ to
each morphism $g\in\sym{C}_1$, such that the following three
conditions hold,
\begin{enumerate}
\item
  For any $2$-morphism $f\colon g\Rightarrow g^\prime$ in $\sym{C}$
  between morphisms $g,g^\prime\colon x\to y$, the diagram
\begin{equation}
\label{eq_bigcan}
\begin{aligned}
\xymatrix{
  Fx\ar[ddd]_{\eta_x}
    \ar@{}[ddd]^(0.85){}="fx"\ar@/^3ex/[rrr]^{Fg}="a"\ar@/_3ex/[rrr]_{Fg^\prime}="b"
  &&&Fy\ar[ddd]^{\eta_y}\ar@{}[ddd]_(0.15){}="fy"\\
  \\
  \\
  \tilde Fx\ar@/^3ex/[rrr]^{\tilde Fg}="c"\ar@/_3ex/[rrr]_{\tilde Fg^\prime}="d"&&&\tilde Fy\\
  \ar@{=>}^{Ff} "a"+<0pt,-2.5ex>;"b"+<0pt,2.5ex>
  \ar@{:>}^{\tilde Ff} "c"+<0pt,-2.5ex>;"d"+<0pt,2.5ex>
  \ar@{} "fy";"c"|(0.3){}="f1"
  \ar@{} "fy";"c"|(0.7){}="f2"
  \ar@{} "b";"fx"|(0.3){}="b1"
  \ar@{} "b";"fx"|(0.7){}="b2"
  \ar@{=>} "f1";"f2"^{\eta_{g^\prime}}
  \ar@{:>} "b1";"b2"_{\eta_g}
}
\end{aligned}
\end{equation}
  $2$-commutes (this means it commutes for $2$-morphisms), \ie\ we
  have the following equality of $2$-morphisms,
\begin{equation}
\label{eq_twonat}
  (Ff\cdot\id^{(2)}_{\eta_y})\circ\eta_{g^\prime}
  = \eta_g\circ(\id^{(2)}_{\eta_x}\cdot \tilde Ff).
\end{equation}
  In~\eqref{eq_bigcan}, the $2$-morphism $\eta_{g^\prime}\colon
  F{g^\prime}\cdot\eta_y\Rightarrow \eta_x\cdot \tilde F{g^\prime}$ is
  located at the front face of the diagram while $\eta_g\colon
  Fg\cdot\eta_y\Rightarrow \eta_x\cdot \tilde Fg$ is at the back.
\item
  For any two composable morphisms $g_1\colon x\to y$ and $g_2\colon
  y\to z$ in $\sym{C}$, the diagram
\begin{equation}
\begin{aligned}
\xymatrix{
  &&&&Fy\ar[drr]^{Fg_2}\ar[ddd]_{\eta_y}
    \ar@{}[ddddllll]|(0.4){}="e"\ar@{}[ddddllll]|(0.55){}="f"\\
  Fx\ar[ddd]_{\eta_x}\ar[urrrr]^{Fg_1}\ar[rrrrrr]^{F(g_1\cdot g_2)}&&&&&&
  Fz\ar[ddd]^{\eta_z}\ar@{}[dddllllll]|(0.12){}="a"\ar@{}[dddllllll]|(0.25){}="b"
    \ar@{}[ddll]|(0.3){}="c"\ar@{}[ddll]|(0.6){}="d"\\
  \\
  &&&&\tilde Fy\ar[drr]^{\tilde Fg_2}\\
  \tilde Fx\ar[urrrr]^{\tilde Fg_1}\ar[rrrrrr]_{\tilde F(g_1\cdot g_2)}&&&&&&\tilde Fz
  \ar@{=>}_{\eta_{g_1\cdot g_2}} "a";"b"
  \ar@{:>}^{\eta_{g_2}} "c";"d"
  \ar@{:>}^{\eta_{g_1}} "e";"f"
}
\end{aligned}
\end{equation}
  which has the equalities $F(g_1\cdot g_2)=Fg_1\cdot Fg_2$ at the top
  and $\tilde F(g_1\cdot g_2)=\tilde Fg_1\cdot\tilde Fg_2$ at the bottom, $2$-commutes,
  \ie\
\begin{equation}
  \eta_{g_1\cdot g_2} =
  (\id^{(2)}_{Fg_1}\cdot\eta_{g_2})\circ(\eta_{g_1}\cdot\id^{(2)}_{\tilde Fg_2}).
\end{equation}
\item
  For each object $x\in\sym{C}_0$, we have $\eta_{\id^{(1)}_x}=\id^{(2)}_{\eta_x}$.
\end{enumerate}
\end{definition}

%
\section{2-form lattice gauge theory}
%
\label{sect_construction}

In the previous section, we have generalized the notions of category,
functor and natural transformation to the next level and we have seen
how one can construct explicit examples using Lie crossed modules. Let
us now generalize the ideas of Section~\ref{sect_ordinary} step by
step in order to obtain a higher level lattice gauge theory,
\emph{lattice $2$-gauge theory}, to be precise. Some key ideas of this
section are taken from~\cite{GrSc01} which deals with $2$-categories
constructed from monoidal categories rather than from Lie $2$-groups.

\subsection{Lattice}

First we have to define a small strict $2$-category which represents
the lattice. It is most convenient to do this for a triangulation
rather than for a cubic lattice\footnote{The material presented here
can be seen as a simplified version of a special case of Street's
construction~\cite{St87}.}.

\begin{definition}
A simplicial $2$-complex $(V,E,F)$ consists of sets $V$
(\emph{vertices}), $E$ (\emph{edges}) and $F$ (\emph{faces}) together
with maps $s\colon E\to V$, $t\colon E\to V$ and
$\del_1,\del_2,\del_3\colon F\to E$ such that $(V,E)$ with $s$ and $t$
forms a directed graph (Example~\ref{ex_graph}). The maps
$\del_1,\del_2,\del_3$ indicate the three edges in the boundary of
each triangular face,
\begin{equation}
\begin{aligned}
\xymatrix{
  v_1\ar[rr]^{\del_1f}&&v_2\ar[dd]^{\del_2f}\\
  \\
  &&v_3\ar[uull]^{\del_3f}="a"
  \ar@{=>}_f [uu]+<-2.5ex,-2.5ex>;"a"+<2.5ex,2.5ex>
}
\end{aligned}
\end{equation}
and are required to satisfy for each $f\in F$,
\begin{equation}
  s(\del_1(f))=t(\del_3(f)),\qquad
  s(\del_2(f))=t(\del_1(f)),\qquad
  s(\del_3(f))=t(\del_2(f)).
\end{equation}
\end{definition}

\begin{example}
\label{ex_complex}
Let $(V,E,F)$ be a simplicial $2$-complex. Then there is a small
strict $2$-category $\sym{C}^{V,E,F}$ defined as follows. The sets
$\sym{C}^{V,E,F}_0$ of objects and $\sym{C}^{V,E,F}_1$ of morphisms
are defined as in Example~\ref{ex_graph}. The maps $s^{(1)}$,
$t^{(1)}$, $\id^{(1)}$ and $\cdot$ are the same as $s,t,\id$ and
$\circ$ in Example~\ref{ex_graph}. This defines a small category which
describes the edges and their compositions. The set
$\sym{C}^{V,E,F}_2$ of $2$-morphisms consists of
\begin{enumerate}
\item
  All faces $f\in F$. We set $s^{(2)}(f)=(\del_1f)\cdot(\del_2f)$ and
  $t^{(2)}(f)={(\del_3f)}^\ast$,
\item
  For each face $f\in F$ another face $f^\ast$ with the double arrow
  reversed,
\begin{equation}
\begin{aligned}
\xymatrix{
  v_1\ar[ddrr]_{{(\del_3f)}^\ast}="a"\ar[rr]^{\del_1f}&&v_2\ar[dd]^{\del_2f}\\
  \\
  &&v_3
  \ar@{<=}_{f^\ast} [uu]+<-2.5ex,-2.5ex>;"a"+<3ex,3ex>
}
\end{aligned}
\end{equation}
  such that $s^{(2)}(f^\ast)={(\del_3f)}^\ast$ and
  $t^{(2)}(f^\ast)=(\del_1f)\cdot (\del_2f)$,
\item
  For each face $f\in F$ another face $\overline{f}$ with all single
  arrows reversed,
\begin{equation}
\begin{aligned}
\xymatrix{
  v_1&&v_2\ar[ll]_{{(\del_1f)}^\ast}\\
  \\
  &&v_3\ar[uu]_{{(\del_2f)}^\ast}\ar[uull]^{\del_3f}="a"
  \ar@{=>}_{\overline f}[uu]+<-2.5ex,-2.5ex>;"a"+<2.5ex,2.5ex>
}
\end{aligned}
\end{equation}
  such that
  $s^{(2)}(\overline{f})={(\del_2f)}^\ast\cdot{(\del_1f)}^\ast$ and
  $t^{(2)}(\overline{f})=\del_3f$,
\item
  For each edge $e\in E$ a $2$-morphism $\id^{(2)}_e$ with
  $s^{(2)}(\id^{(2)}_e)=e=t^{(2)}(\id^{(2)}_e)$,
\item
  All formal horizontal $(\cdot)$ and vertical $(\circ)$ compositions
  of faces, subject to the following relations,
\begin{enumerate}
\item
  $f\circ f^\ast=\id^{(2)}_{s^{(2)}(f)}$ and $f^\ast\circ
  f=\id^{(2)}_{t^{(2)}(f)}$,
\item
  $\id^{(2)}_{s^{(2)}(f)}\circ f=f=f\circ\id^{(2)}_{t^{(2)}(f)}$,
\item
  $f\circ\overline{f}=\id^{(2)}(\id^{(1)}(s^{(1)}(s^{(2)}(f))))$ and
  $\overline{f}\circ f=\id^{(2)}(\id^{(1)}(t^{(1)}(s^{(2)}(f))))$,
\item
  $\id^{(2)}(\id^{(1)}(s^{(1)}(s^{(2)}(f))))\cdot f = f = 
    f\cdot \id^{(2)}(\id^{(1)}(t^{(1)}(t^{(2)}(f))))$,
\item
  $\id^{(2)}_{e_1\cdot e_2}=\id^{(2)}_{e_1}\cdot\id^{(2)}_{e_2}$ for
  all composable edges $e_1,e_2\in E$,
\item
  the exchange law, $(f_1\cdot f_2)\circ(f_1^\prime\cdot
  f_2^\prime)=(f_1\circ f_1^\prime)\cdot(f_2\circ f_2^\prime)$,
  whenever faces $f_1,f_1^\prime,f_2,f_2^\prime$ are composable.
\end{enumerate}
\end{enumerate}
\end{example}

\subsection{Configurations}

Let us now study the configurations of our generalized lattice gauge
theory. By analogy with Section~\ref{sect_ordinary}, these are the
$2$-functors from the generalized lattice $\sym{C}^{V,E,F}$ to the
generalized gauge group $\sym{G}^{G,H}$.

\begin{example}
A strict $2$-functor $F\colon \sym{C}^{V,E,F}\to\sym{G}^{G,H}$ is a
triple of maps
\begin{eqnarray}
  &F_0&\colon V \to \{\ast\},\\
  &F_1&\colon E \to G,\\
  &F_2&\colon F \to H\rtimes G,
\end{eqnarray}
\ie\ the edges are coloured by group elements of $G$ while the
triangular faces are coloured by elements of $H\rtimes G$.
\end{example}

Let us assume there is such a strict $2$-functor which is used to
label the triangle $(1,2,3)$,
\begin{equation}
\label{eq_labeled}
\begin{aligned}
\xymatrix{
  1\ar[ddrr]_{g_{13}}="a"\ar[rr]^{g_{12}}&&2\ar[dd]^{g_{23}}\\
  \\
  &&3
  \ar@{=>}_{f_{123}} [uu]+<-2.5ex,-2.5ex>;"a"+<2.5ex,2.5ex>
}
\end{aligned}
\end{equation}
We denote by $f_{123}:=F_2(f)\in H\rtimes G$ the group element
associated with the triangle and by $g_{12}:=\del_1f$,
$g_{23}:=\del_2f$ and $g_{13}:={(\del_3f)}^{-1}$ the elements of $G$
associated with the edges. Then we can derive a number of useful
properties of such a labeled triangle.

\begin{enumerate}
\item
  Write $f_{123}=(h,g)\in H\rtimes G$. We know that
  $s^{(2)}(f)=g_{12}\cdot g_{23}$ and $t^{(2)}(f)=g_{13}$ so that
  $t(h)=\alignidx{g_{13}g_{23}^{-1}g_{12}^{-1}}$ which is just the holonomy
  around the triangle.
\item
  We can horizontally compose the $2$-morphism $f_{123}$ with
  identities,
\begin{equation}
  {\hat f}_{123}:=\id^{(2)}_{g_{12}^{-1}}\cdot f_{123}\cdot
  \id^{(2)}_{g_{13}^{-1}}\colon \alignidx{g_{23}\cdot g_{13}^{-1}}\Rightarrow
  g_{12}^{-1},
\end{equation}
  which can be visualized as follows,
\begin{equation}
\begin{aligned}
\xymatrix{
  1&&2\ar[ll]_{g_{12}^{-1}}="a"\ar[dd]^{g_{23}}\\
  \\
  &&3\ar[uull]^{g_{13}^{-1}}
  \ar@{=>}_{{\hat f}_{123}} []+<-3ex,6ex>;"a"+<2ex,-4ex>
}
\end{aligned}
\end{equation}
  We have thus obtained another triangle on which the arrow for the
  $2$-morphism has been `rotated'. On our generalized lattice
  $\sym{C}^{V,E,F}$ (Example~\ref{ex_complex}), this new `rotated'
  triangle is considered different from the original one. At first
  sight, it seems that triangles proliferate if we `rotate' them in
  this way. In our generalized gauge group $\sym{G}^{G,H}$, one can
  show, however, that `triple rotation' does not change the associated
  $2$-morphism because
\begin{equation}
  \id^{(2)}_{g_{13}}\cdot \id^{(2)}_{g_{23}^{-1}}\cdot \id^{(2)}_{g_{12}^{-1}}\cdot f_{123}
  \cdot \id^{(2)}_{g_{13}^{-1}}\cdot \id^{(2)}_{g_{12}}\cdot \id^{(2)}_{g_{23}} = f_{123}.
\end{equation}
\item
  From the conditions on the identities in Example~\ref{ex_complex}
  and the properties of the strict $2$-functor, we know that for each
  vertex $v\in V$, the identity edge $\id^{(1)}_v$ is mapped to the
  unit in $G$,
\begin{equation}
  F_1\id^{(1)}_v=\id^{(1)}_\ast=1\in G,
\end{equation}
  while for any edge $e$, labeled by $g:=F_1e$, the unit $\id^{(2)}_e$
  is mapped to,
\begin{equation}
  F_2\id^{(2)}_e=\id^{(2)}_g=(1,g)\in H\rtimes G.
\end{equation}
  These conditions imply that for each $2$-morphism $f_{123}$, the
  $2$-morphism $f^\ast_{123}$ (Example~\ref{ex_complex}) is given by
  its inverse with respect to vertical composition,
\begin{equation}
  f_{123}^\ast = {(h,g)}^\ast = (h^{-1},t(h)g),
\end{equation}
  while the other $2$-morphism $\overline{f}_{123}$ is the inverse
  with respect to horizontal composition,
\begin{equation}
  \overline{f}_{123} = {(h,g)}^{-1} = (\alpha(g^{-1})[h^{-1}],g^{-1}).
\end{equation}
\item
  Finally, these two ways of reversing the orientation of triangles
  are related in the following way,
\begin{equation}
  \id^{(2)}_{g_{23}}\cdot\overline{f^\ast}_{123}\cdot\id^{(2)}_{g_{13}}
  = \id^{(2)}_{g_{12}^{-1}}\cdot f.
\end{equation}
  A careful analysis shows that this is exactly what one expects from
  the geometry of the triangle if these operations are combined.
\end{enumerate}

Starting from the labeled triangle~\eqref{eq_labeled}, we can obtain
other $2$-morphisms for the same triangle by `rotating' or reversing
the orientation using either $f\mapsto f^\ast$ or
$f\mapsto\overline{f}$. The relations listed above make sure that one
obtains only six distinct $2$-morphisms by combining these
operations. The $2$-groupoid $\sym{G}^{G,H}$ has therefore all the
properties which one expects from the combinatorics of the
triangle~\eqref{eq_labeled}. While edges come in two different
orientations, there are six different versions of each triangle.

\subsection{Local gauge transformations}

Given two configurations of our generalized lattice gauge theory,
represented by strict $2$-functors from $\sym{C}^{V,E,F}$ to
$\sym{G}^{G,H}$, the analogy with Section~\ref{sect_ordinary} suggests
that the local gauge transformations are given by pseudo-natural
transformations. In our situation, this reads as follows.

\begin{example}
\label{ex_local2}
Let $F,\tilde F\colon\sym{C}^{V,E,F}\to\sym{G}^{G,H}$ be parallel
strict $2$-functors. A pseudo-natural transformation $\eta\colon
F\Rightarrow\tilde F$ is a pair of maps
\begin{alignat}{2}
  \eta&\colon V\to G,&\quad& v\mapsto \eta_v,\\
  \eta&\colon E\to H\rtimes G,&\quad& e\mapsto 
    \eta_e\colon F_1e\cdot\eta_{t(e)}\Rightarrow \eta_{s(e)}\cdot\tilde F_1e,
\end{alignat}
visualized for an edge $e\colon v\to w$ by
\begin{equation}
\begin{aligned}
\xymatrix{
  \ast\ar[dd]_{\eta_v}\ar[rr]^{Fe}&&\ast\ar[dd]^{\eta_w}\ar@{}[ddll]|(0.5){}="b"\ar@{}[ddll]|(0.2){}="a"\\
  \\
  \ast\ar[rr]_{\tilde Fe}&&\ast
  \ar@{=>}^{\eta_e} "a";"b"
}
\end{aligned}
\end{equation}
\end{example}

The special case in which $\eta_e=\id^{(2)}_e=(1,1)$, corresponds to
an ordinary local gauge transformation because source and target of
$\eta_e$ agree and therefore the diagram commutes for
morphisms. Compare this with Example~\ref{ex_localgauge}. The
appearance of a non-trivial $2$-morphism $\eta_e$ can be viewed as a
way of parameterizing how `non-commuting' the diagram is. This is the
way in which the local gauge symmetry is generalized here. We call the
pseudo-natural transformations of the above example the \emph{local
$2$-gauge transformations}.

Let us now visualize how a generic local $2$-gauge transformation acts
on the labeled triangle~\eqref{eq_labeled},
\begin{equation}
\label{eq_prism}
\begin{aligned}
\xymatrix{
  &&&&\ast\ar[drr]^{g_{23}}\ar[ddd]_{\eta_2}
    \ar@{}[ddddllll]|(0.4){}="e"\ar@{}[ddddllll]|(0.55){}="f"\\
  \ast\ar[ddd]_{\eta_1}\ar[urrrr]^{g_{12}}\ar[rrrrrr]_{g_{13}}="g"&&&&&&
  \ast\ar[ddd]^{\eta_3}\ar@{}[dddllllll]|(0.12){}="a"\ar@{}[dddllllll]|(0.25){}="b"
    \ar@{}[ddll]|(0.3){}="c"\ar@{}[ddll]|(0.6){}="d"\\
  \\
  &&&&\ast\ar[drr]^{{\tilde g}_{23}}\\
  \ast\ar[urrrr]^{{\tilde g}_{12}}\ar[rrrrrr]_{{\tilde g}_{13}}="h"&&&&&&\ast
  \ar@{=>}_{\eta_{13}} "a";"b"
  \ar@{:>}^{\eta_{23}} "c";"d"
  \ar@{:>}^{\eta_{12}} "e";"f"
  \ar@{=>}_{f_{123}} [uuuull]+<-1.5ex,-2.5ex>;"g"+<2ex,2.5ex>
  \ar@{:>}_{{\tilde f}_{123}} [ull]+<-1.5ex,-2.5ex>;"h"+<2ex,2.5ex>
}
\end{aligned}
\end{equation}
Here we have denoted by $f_{123}$, $g_{12}$, \etc\ the face and edges
labeled by the strict $2$-functor $F$ and by ${\tilde f}_{123}$,
${\tilde g}_{12}$, \etc\ the face and edges labeled by $\tilde F$. The
three squares in~\eqref{eq_prism} are labeled by the $2$-morphisms
$\eta_{ij}:=\eta_{e_{ij}}\colon g_{ij}\cdot \eta_j\Rightarrow
\eta_i\cdot\tilde g_{ij}$. By condition~(1)--(3) of
Definition~\ref{def_pseudonat}, the diagram~\eqref{eq_prism}
$2$-commutes. Therefore we can calculate ${\tilde f}_{123}$ from,
\begin{equation}
\label{eq_twogauge}
  {\tilde f}_{123} = \id^{(2)}_{\eta_1^{-1}}\cdot\Bigl(
    (\eta_{12}^\ast\cdot \id^{(2)}_{\tilde g_{23}})\circ
    (\id^{(2)}_{g_{12}}\cdot\eta_{23}^\ast)\circ f_{123}\circ\eta_{13}\Bigr).
\end{equation}
Observe that the right hand side involves both $g_{ij}$ and ${\tilde
g}_{ij}$.

\subsection{Gauge invariant expressions}

In standard lattice gauge theory, gauge invariant quantities can be
constructed from the holonomy around closed loops, namely by
evaluating a group character. In order to find expressions that are
invariant under local $2$-gauge transformations, we consider the
vertical composition of $2$-morphisms over a closed surface, in the
simplest case a tetrahedron. While the holonomy around the loop was
based at a point, our vertical composition is now based at an edge of
the tetrahedron.

\begin{figure}[t]
\begin{center}
\input{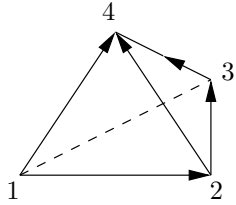}
\end{center}
\caption{\label{fig_tetrahedron}%
A tetrahedron with vertices labeled $1,2,3,4$. Each triangle
$(i,j,k)$, $i<j<k$, is coloured as in~\eqref{eq_labeled}.}
\end{figure}

Figure~\ref{fig_tetrahedron} shows a tetrahedron $(1,2,3,4)$. We label
its triangles $(i,j,k)$, $i<j<k$, as in~\eqref{eq_labeled}. A
$2$-morphism around the surface of the tetrahedron can be calculated
as follows.

\begin{definition}
\label{def_surface}
Let $F\colon\sym{C}^{V,E,F}\to\sym{G}^{G,H}$ be a strict
$2$-functor. For each tetrahedron $(1,2,3,4)$, the \emph{$2$-holonomy}
is the $2$-morphism $\Phi_{1234}\colon g_{14}\Rightarrow g_{14}$ in
$\sym{G}^{G,H}$, given by,
\begin{equation}
  \Phi_{1234}:=f^\ast_{124}
               \circ (\id^{(2)}_{g_{12}}\cdot f^\ast_{234})
               \circ (f_{123}\cdot\id^{(2)}_{g_{34}})
	       \circ f_{134}.
\end{equation}
\end{definition}

Since any $2$-morphism associated with a closed surface has the same
source and target, $\Phi_{1234}=(h_{1234},g_{14})$ is characterized by
an element $h_{1234}\in\ker t\unlhd H$. It is therefore sufficient to
define our gauge invariant actions as functions on $\ker t$. 

Notice that we have labeled the $1$- and $2$-cells (edges and faces)
of the lattice with data from the $2$-groupoid $\sym{G}^{G,H}$, but
that the $2$-action is defined one level higher, namely for the
$3$-cells (tetrahedra). So far we have not formally defined the
notion of a tetrahedron. For our purposes it is sufficient that the
preceding definition can be used whenever we have a collection of four
triangles whose edges match as shown in Figure~\ref{fig_tetrahedron}.

In standard lattice gauge theory, the action is a character of the
holonomy around a loop. In our generalized setting, it turns out that,
for our $2$-holonomy, we need the following two invariances in order
to obtain a locally $2$-gauge invariant action.

\begin{definition}
Let $(G,H,t,\alpha)$ be a Lie crossed module. A $2$-action is a
function $S\colon H\rtimes G\to\R$ which is the composition
$S(\Phi)=s_0(\pi(\Phi))$ of a function $s_0\colon \ker t\to\R$ with
the projection $\pi\colon H\rtimes G\to H$. We define the function
$s_0$ only on $\ker t\unlhd H$. It is required to be a class function
of $H$, \ie\
\begin{equation}
  s_0(h^\prime h{h^\prime}^{-1})=s_0(h),
\end{equation}
for all $h\in \ker t$, $h^\prime\in H$, and to be constant on the
orbits of $G$, \ie\
\begin{equation}
  s_0(\alpha(g)[h]) = s_0(h),
\end{equation}
for all $h\in \ker t$ and $g\in G$.
\end{definition}

We present examples and discuss possible physical applications in
Section~\ref{sect_examples} below. We have required the two
invariances for the following purpose.

\begin{lemma}
\label{lemma_action}
Let $(G,H,t,\alpha)$ be a Lie crossed module and $S\colon H\rtimes
G\to \R$ be a $2$-action. Let $f=(h,g)\in H\rtimes G$ be any $2$-morphism
$f\colon g\Rightarrow g$, \ie\ $h\in\ker t$. Then $S(f)$ is invariant
under horizontal composition with identities because for any
$\id^{(2)}_{\tilde g}=(1,\tilde g)$, $\tilde g\in G$,
\begin{equation}
  (h,g)\cdot (1,\tilde g) = (h,g\tilde g)\qquad\mbox{and}\qquad
  (1,\tilde g)\cdot (h,g) = (\alpha(\tilde g)[h],\tilde g g),
\end{equation}
and we have $s_0(\alpha(\tilde g)[h])=s_0(h)$. Furthermore, $S(f)$ is
invariant under vertical conjugation because for any $(\tilde h,\tilde
g)\in H\rtimes G$,
\begin{equation}
  {(\tilde h,\tilde g)}^\ast\circ(h,g)\circ(\tilde h,\tilde g) = ({\tilde h}^{-1}h\tilde h,\tilde g),
\end{equation}
and $s_0({\tilde h}^{-1}h\tilde h)=s_0(h)$.
\end{lemma}

\begin{figure}[t]
\begin{center}
\input{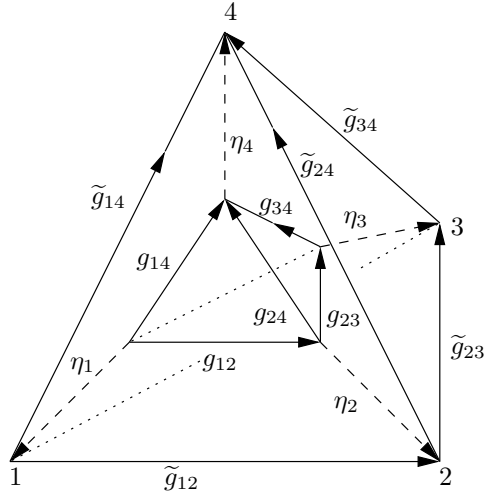}
\end{center}
\caption{\label{fig_commutative}%
The inner tetrahedron is labeled by a strict $2$-functor
$F\colon\sym{C}^{V,E,F}\to \sym{G}^{G,H}$, the outer one by some
strict $2$-functor $\tilde F$ (notation as in~\eqref{eq_labeled}). Both
$2$-functors are related by a pseudo-natural transformation
$\eta\colon F\Rightarrow\tilde F$. For simplicity, we have not drawn the
double arrows on the faces.}
\end{figure}

\begin{theorem}
\label{thm_gaugeinv}
Let $(V,E,F)$ be a simplicial $2$-complex and $(G,H,t,\alpha)$ be a
Lie crossed module. Let $F,\tilde
F\colon\sym{C}^{V,E,F}\to\sym{G}^{G,H}$ be parallel strict
$2$-functors and $S\colon H\rtimes G\to\R$ be some $2$-action. If
there exists a pseudo-natural transformation $\eta\colon
F\Rightarrow\tilde F$, then the $2$-action, evaluated on any
tetrahedron $(1,2,3,4)$ in $(V,E,F)$ agrees for $F$ and $\tilde F$,
\ie
\begin{equation}
  S(\Phi_{1234}) = S({\tilde \Phi}_{1234}),
\end{equation}
where $\Phi_{1234}$ is the $2$-holonomy of
Definition~\ref{def_surface} using the strict $2$-functor $F$ and
${\tilde\Phi}_{1234}$ using $\tilde F$.
\end{theorem}

\begin{proof}
According to Definition~\ref{def_surface},
\begin{equation}
  {\tilde \Phi}_{1234}
  = {\tilde f}^\ast_{124}
    \circ(\id^{(2)}_{{\tilde g}_{12}}\cdot {\tilde f}^\ast_{234})
    \circ({\tilde f}_{123}\cdot\id^{(2)}_{{\tilde g}_{34}})
    \circ{\tilde f}_{134},
\end{equation}
and
\begin{equation}
  \Phi_{1234}
  = f^\ast_{124}
    \circ(\id^{(2)}_{g_{12}}\cdot f^\ast_{234})
    \circ(f_{123}\cdot\id^{(2)}_{g_{34}})
    \circ f_{134}.
\end{equation}
The two coloured tetrahedra corresponding to $\Phi_{1234}$ and
${\tilde \Phi}_{1234}$ with the pseudo-natural transformation $\eta$
are shown in Figure~\ref{fig_commutative}. On the faces of this
diagram, there are $2$-morphisms from the strict $2$-functors $F$ and
$\tilde F$ and also from the pseudo-natural transformation $\eta$
which we have suppressed in order to keep the drawing transparent. The
four prisms attached to the triangular faces of the inner tetrahedron
are of the form~\eqref{eq_prism} and therefore $2$-commute. We read
off from the picture that,
\begin{equation}
  {\tilde\Phi}_{1234}=\id^{(2)}_{\eta_1^{-1}}\cdot\Bigl(
  \eta_{14}^\ast
  \circ(\Phi_{1234}\cdot\id^{(2)}_{\eta_4})
  \circ\eta_{14}\Bigr),
\end{equation}
so that $\Phi_{1234}$ and ${\tilde \Phi}_{1234}$ are related by
horizontal composition with identities and by vertical
conjugation. Therefore, the value of $S$ agrees according to
Lemma~\ref{lemma_action}.
\end{proof}

\begin{remark}
It can be shown that the edge, here $e_{14}$, on which the
$2$-morphism $\Phi_{1234}$ is based, does not matter. By horizontal
composition with identity $2$-morphisms we can obtain an analogous
$2$-morphism based on any other edge which yields the same value of
the action.

Of course, we can calculate locally $2$-gauge invariant expressions
for any closed surface by calculating appropriate compositions of the
$2$-morphisms. Pasting theorems, see, for example~\cite{Po90},
guarantee that the $2$-holonomy is well defined and independent of
choices.

The gauge invariant expressions of the standard formulation of lattice
gauge theory are in general not invariant under local $2$-gauge
symmetry transformations unless all $2$-morphisms $\eta_e$ associated
with the edges have the same source and target.
\end{remark}

\begin{remark}
As $\ker t$ corresponds to the set of all $2$-morphisms whose source
and target coincide, the Eckmann--Hilton argument implies that $\ker
t$ is always Abelian (in fact, it is always contained in the centre of
$H$). Even though both $G$ and $H$ can be non-Abelian and there is a
non-trivial interplay between the two via $t$ and $\alpha$, the
quantities on which the $2$-actions depend, are therefore always
Abelian.
\end{remark}

This completes the construction of our generalization of lattice gauge
theory. The generalized lattice and the generalized gauge group are
both described by $2$-categories. The configurations are given by
strict $2$-functors, the local $2$-gauge symmetries by pseudo-natural
transformations. In the last step, we have found actions that are
invariant under this generalized local gauge symmetry.

\subsection{Partition function}
\label{sect_partition}

It is now straightforward to write down a path integral for a quantum
version of our lattice $2$-gauge theory. Integrate over $G$ for each
edge and over $H\rtimes G$ for each triangle, subject to the condition
($\delta$-functions on $G$) that source and target of the
$2$-morphisms associated to the faces match. The integrand is the
product over $w(\Phi_{jk\ell m}):=\exp(-S(\Phi_{jk\ell m}))$ or
$\exp(iS(\Phi_{jk\ell m}))$ for all tetrahedra $(jk\ell m)$ depending
on whether Euclidean or real time is used. Here $S$ denotes some
$2$-action. The partition function therefore reads,
\begin{eqnarray}
  Z &=& \Bigl(\prod_{e\in E}\int_G\,dg_e\Bigr)
    \Bigl(\prod_{t\in F}\int_{H\rtimes G}\,df_t\Bigr)\nn\\
    &&\times\Bigl(\prod_{t\in F}\delta_G(s^{(2)}(f_t)\cdot{(g_{\del_1t}g_{\del_2t})}^{-1})
      \delta_G(t^{(2)}(f_t)\cdot g_{\del_3t})\Bigr)
    \Bigl(\prod_{\sigma\in T}w(\Phi_\sigma)\Bigr).
\end{eqnarray}
Here we have denoted the triangles by $t\in F$ and the tetrahedra by
$\sigma\in T$. We already know some observables of this theory, namely
the expectation values of the $2$-gauge invariant expressions
constructed in Theorem~\ref{thm_gaugeinv}.

For any configuration given by $g_e\in G$ for each edge $e\in E$ and
$f_t\in H\rtimes G$ for each triangle $t\in F$, we obtain a locally
$2$-gauge equivalent configuration by applying the pseudo-natural
transformation~\eqref{eq_twogauge}.

%
\section{Examples and physical applications}
%
\label{sect_examples}

In this section, we come back to some examples of strict Lie
$2$-groups (Example~\ref{ex_examples}), illustrate what the admissible
gauge invariant $2$-actions are and sketch possible applications.

\begin{example}
The trivial $2$-group (see Example~\ref{ex_examples}(1)). In this case
the assignment of variables reduces to conventional lattice gauge
theory. There are no labels at the faces, and the local gauge
transformations (Example~\ref{ex_local2}) reduce to the ordinary ones
(Example~\ref{ex_localgauge}). As there are no variables associated
with the faces, $2$-actions are useless in this case.
\end{example}

\begin{example}
The purely Abelian $2$-group (see Example~\ref{ex_examples}(2)). In
this case, the edges are not labeled while the faces are labeled with
elements of some Abelian group $H$. We have $\ker t=H$. If we choose
$H=\U(1)$, we recover $2$-form electrodynamics. Possible $2$-actions
are real characters of $H$, \ie\ for $G=\U(1)$ they are of the form
$s_0(e^{i\phi})=\cos (k\phi)$ for some $k\in\Z$, and our model agrees
with that of~\cite{Sa77,Sa80} for $p=2$ except that it is defined on
triangulations rather than on hyper-cubic lattices.
\end{example}

\begin{example}
The Euclidean $2$-groups (see Example~\ref{ex_examples}(4)). Again we
have $\ker t=H$. The allowed $2$-actions are functions of the
$\SO(V,\eta)$-invariant norm, \ie\ of the form $s_0(h)=f(\eta(h,h))$
for some function $f\colon\R\to\R$. This resembles the expansors used
in~\cite{CrYe03a} in the case of the Poincar\'e $2$-group.

Observe that this example is a non-trivial generalization of $2$-form
electrodynamics even though $H$ is Abelian. There is a non-trivial
interplay with $G$ which requires the $2$-action to be constant on the
orbits of $G$ on $H$. For the Poincar\'e $2$-group, this example
suggests the following more ambitious conjecture.
\end{example}

\begin{conjecture}
Lattice $2$-gauge theory using the Poincar\'e $2$-group in four
dimensions with a suitable action is equivalent to the refined
Barrett--Crane--Yetter state sum proposed in~\cite{CrYe03a} in the
same fashion as standard lattice gauge theory with $\delta$-function
Boltzmann weight is related to the Ooguri model, see, for
example~\cite{Re94,Ba99}.

Lattice $2$-gauge theory should therefore be formulated on the
$2$-complex dual to the triangulation used in the state sum model so
that we can apply a suitable harmonic analysis and obtain sums over
representations of $H$ for the edges and sums over representations of
$G$ for the faces of the triangulation. This way it might be possible
to obtain further conditions on the measure used in the state sum
model and to relate it to a classical action with constraints at the
classical level.
\end{conjecture}

\begin{example}
The automorphism $2$-group of a simple compact Lie group $H$ (see
Example~\ref{ex_examples}(5)). In this case $\ker t=Z(H)$ is the
centre of $H$ and any function $s_0\colon Z(H)\to\R$ gives rise to a
$2$-action. This example can be understood as a lattice model of the
connections on non-Abelian gerbes. We notice some coincidences which
suggest the following conjecture.
\end{example}

\begin{conjecture}
Lattice $2$-gauge theory with the automorphism $2$-group for
$H=\SU(3)$, \ie\ $G=\SU(3)/\Z_3$ and $\ker t=\Z_3$, describes some
aspects of the collective phenomena of strongly coupled pure QCD. In
fact, the $2$-gauge invariant expressions of
Theorem~\ref{thm_gaugeinv} resemble the observables that detect centre
monopoles and vortices which seem to play a key role in the
understanding of the confinement of static quarks in pure lattice QCD.
\end{conjecture}

%
\section{Discussion and outlook}
%
\label{sect_conclude}

\subsection{Technical questions}

At the technical level, there are a number of natural questions to
ask. What is the most general gauge invariant expression? In standard
lattice gauge theory, these are \emph{spin networks}~\cite{RoSm95b},
generalizations of Wilson loops that include branchings of the lines
with intertwiners of the gauge group at the branching points. In order
to fully understand the possible spin networks, one has to study the
representation category of the gauge group. In our lattice $2$-gauge
theory, the expressions $S(\Phi_{1234})$ of Theorem~\ref{thm_gaugeinv}
are the analogues of Wilson loops. The most general gauge invariant
expressions will be given by coloured branched surfaces, \ie\ by some
sort of \emph{spin foams}~\cite{Ba98}. In order to understand these
spin foams, we have to study the representation $2$-category of a Lie
$2$-group. Work on the representation theory of $2$-groups is in
progress~\cite{CrYe03b,BaMa03}. It is desirable to find a gauge
invariant expression which combines both holonomies of morphisms
around loops and $2$-holonomies around closed surfaces so that this
expression reduces to the Wilson action if we choose the trivial
$2$-group. At the moment we have either the old observables
(characters of holonomies) which are not $2$-gauge invariant or very
special new ones ($2$-actions of $2$-holonomies) which disappear for
the trivial $2$-group.

A related question is that of a maximal gauge fixing. In standard
lattice gauge theory, one can gauge fix all edges of a spanning tree
to be labeled by the group unit. In lattice $2$-gauge theory, the
gauge fixing will make use of suitable surfaces.

A further technical observation is that for $2$-categories, there is
one more natural type of maps besides functors and pseudo-natural
transformations. These are called \emph{modifications} and relate two
natural transformations. 

\begin{definition}
Let $\sym{C}$, $\sym{C}^\prime$ be small strict $2$-categories,
$F,\tilde F\colon\sym{C}\to\sym{C}^\prime$ be strict $2$-functors and
$\eta,\theta\colon F\Rightarrow \tilde F$ be pseudo-natural
transformations. A \emph{quasi-modification}
$\xymatrix@1{\mu\colon\eta\ar@3{->}[r]&\theta}$ is a map
$\sym{C}_0\to\sym{C}^\prime_2$ assigning to each object
$x\in\sym{C}_0$ a $2$-morphism $\mu_x\colon\eta_x\Rightarrow\theta_x$
in $\sym{C}^\prime$ such that,
\begin{equation}
  (Ff\cdot\mu_y)\circ\theta_{g^\prime} = \eta_{g}\circ(\mu_x\cdot\tilde Ff)
\end{equation}
holds for each $2$-morphism $f\colon g\Rightarrow g^\prime$ in $\sym{C}$
where $g,g^\prime\colon x\to y$ are morphisms in $\sym{C}$. This is
illustrated by the following diagram,
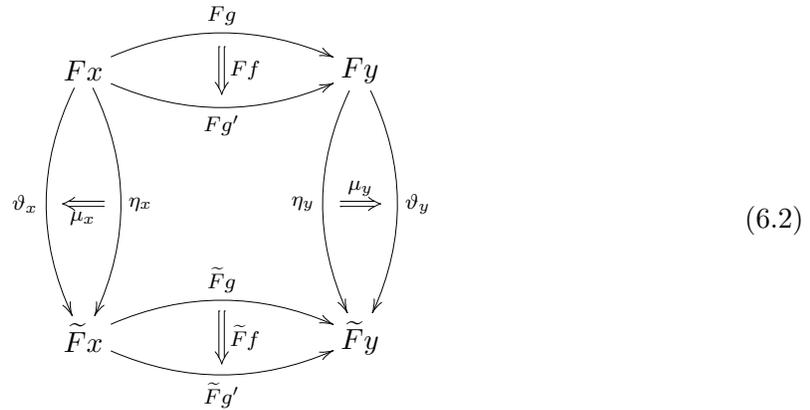
\begin{equation}
\begin{aligned}
\xymatrix{
  Fx\ar@/^3ex/[ddd]^{\eta_x}="e"\ar@/_3ex/[ddd]_{\theta_x}="f"
    \ar@/^3ex/[rrr]^{Fg}="a"\ar@/_3ex/[rrr]_{Fg^\prime}="b"
  &&&Fy\ar@/_3ex/[ddd]_{\eta_y}="g"\ar@/^3ex/[ddd]^{\theta_y}="h"\\
  \\
  \\
  \tilde Fx\ar@/^3ex/[rrr]^{\tilde Fg}="c"\ar@/_3ex/[rrr]_{\tilde Fg^\prime}="d"&&&\tilde Fy\\
  \ar@{=>}^{Ff} "a"+<0pt,-2.5ex>;"b"+<0pt,2.5ex>
  \ar@{=>}^{\tilde Ff} "c"+<0pt,-2.5ex>;"d"+<0pt,2.5ex>
  \ar@{=>}^{\mu_x} "e"+<-3ex,0pt>;"f"+<3ex,0pt>
  \ar@{=>}^{\mu_y} "g"+<3ex,0pt>;"h"+<-3ex,0pt>
}
\end{aligned}
\end{equation}
We have suppressed the double arrows for the following $2$-morphisms
in order to keep the diagram simple,
\begin{alignat}{2}
  \eta_{g}         &\colon F_{g}\cdot\eta_y   
    &\Rightarrow& \eta_x\cdot\tilde F_{g},\\
  \eta_{g^\prime}  &\colon F_{g^\prime}\cdot\eta_y   
    &\Rightarrow& \eta_x\cdot\tilde F_{g^\prime},\\
  \theta_{g}       &\colon F_{g}\cdot\theta_y 
    &\Rightarrow& \theta_x\cdot\tilde F_{g},\\
  \theta_{g^\prime}&\colon F_{g^\prime}\cdot\theta_y 
    &\Rightarrow& \theta_x\cdot\tilde F_{g^\prime}.
\end{alignat}
\end{definition}

It is open what modifications mean physically. Are some gauge
equivalent configurations `more equal' than others? An explanation why
we are not forced to use this additional structural level may be the
fact that our action is a map into the real numbers as opposed to a
map into some $2$-category. The requirement to use real numbers is, of
course, imposed by the physical framework, but the $2$-categorical
treatment might indicate that one should try to categorify the action
or the path integral.

Given a path integral formulation of lattice $2$-gauge theory as
sketched in Section~\ref{sect_partition}, we have the
\emph{connection picture} of this theory which is given in terms of
continuous variables. It is known that for lattice $BF$-theory (see,
for example~\cite{Re94,Ba99}) and also for standard lattice
Yang--Mills theory~\cite{OePf01,PfOe02}, there is an equivalent
\emph{dual} formulation provided by the \emph{representation picture}
which is, in the case of Yang--Mills theory, the corresponding
strong-weak dual theory. The dual formulation of standard lattice
gauge theory is a spin foam model. How does the dual formulation of
lattice $2$-gauge theory look like?

It is already obvious that we have to understand the representation
$2$-category of the gauge $2$-group in order to formulate such a
model. If our lattice $2$-gauge theory lives on the two-complex dual
to some triangulation of a given four-manifold, the dual theory will
involve sums over suitable vector spaces for all edges and for all
faces of the original triangulation. As already mentioned, this is
precisely the structure of the refined Barrett--Crane--Yetter state
sum model of quantum gravity as proposed in~\cite{CrYe03a}. An
interesting project is therefore to perform the harmonic expansion
(better: $2$-harmonic expansion) of the Lie $2$-group valued variables
and, for generic $2$-action, to generalize the transformation
of~\cite{OePf01,PfOe02} to lattice $2$-gauge theory. We emphasize that
the dual theory will involve one higher level of the given simplicial
complex because the $2$-action is associated with the $3$-simplices.

Thinking about the representation picture of lattice $2$-gauge theory,
the entire program of the lattice gauge and state sum models is worth
reconsidering: topological models, diagrammatic techniques
generalizing the chain mail~\cite{Ro95} and generalizing the ribbon
diagrammatics of~\cite{Pf01,Oe01,GiOe02}. What is a suitable
`non-commutative' structure which generalizes the ribbon diagrams
of~\cite{Pf01,Oe01,GiOe02} and maybe weakens the axioms of the
$2$-category used? For weak versions of $2$-groups, see~\cite{La02}. A
general framework for state sum invariants of four-manifolds is
provided by Mackaay's construction~\cite{Ma99}.

\subsection{A hierarchy of theories}

One of the key ideas of higher category theory is that there is a
hierarchy of structures (sets, categories, $2$-categories,
$\ldots$). In this hierarchy, we call standard lattice gauge theory a
\emph{$1$-gauge theory} and the model constructed in the present
article a \emph{$2$-gauge theory}. Lattice $3$-gauge theory is beyond
the scope of this study, but we can still consider lattice $0$-gauge
theory in order to learn more about the hierarchy of models.

Consider the construction of Section~\ref{sect_ordinary}. How does it
collapse in the case of a $0$-gauge theory? The lattice is a
$0$-category, \ie\ just a set, a collection of points without
structure. Similarly the set of labels (`$0$-gauge group'). A
configuration is then a $0$-functor which is just a map from the set
of points to the $0$-gauge group. There is no notion of natural
equivalence and therefore no local symmetry. Such a model resembles a
lattice spin model. These models include the lattice versions of
non-linear sigma models. In total, we have the following information
on the hierarchy of lattice $n$-gauge theories.

$n=0$. No local symmetry. The variables are associated with the
vertices and the action terms with the edges. Gauge `invariant'
quantities are arbitrary functions. For models with a certain global
symmetry, the strong coupling expansion~\cite{Pf02} is an expansion in
terms of spin networks (coloured graphs) and the dual formulation is a
spin network model.

$n=1$. Standard lattice gauge theory. The variables are associated
with the edges and the action terms with the faces. Gauge invariant
quantities are spin networks. The strong coupling
expansion~\cite{OePf01,PfOe02} is an expansion in terms of spin foams
(coloured $2$-complexes) and the dual model is a spin foam model.

$n=2$. The model constructed in the present article. The generic gauge
invariant quantities are certain spin foams. From the assignment of
variables one can already see that the strong coupling expansion will
lead to coloured $3$-complexes.

It seems that we can minimally couple the model at level $n$ with the
model at $n+1$. The classic example at $n=0$ is the Abelian Higgs
model with frozen radial mode~\cite{EiSa78}, but this construction can
be extended to non-Abelian symmetry groups as well~\cite{Pf02}. The
next more general step would therefore be to couple a $2$-gauge theory
to standard lattice gauge theory.

\subsection{General comments}

It is an interesting question whether one can construct a Topological
Quantum Field Theory (TQFT) from our $2$-gauge theory. We should
therefore require the higher level analogue of the flatness condition
in order to obtain topological invariants. This means we should
restrict the partition function of our quantum theory to those
configurations for which the $2$-holonomy $\Phi_\sigma$ vanishes at
every tetrahedron $\sigma$. This is the \emph{zero $2$-curvature
condition}.

For certain finite crossed modules this TQFT was constructed by
Yetter~\cite{Ye93} using the language of categorical groups, \ie\
group objects in the category of groupoids. In the topological case
one can choose an arbitrary triangulation as the lattice and then
construct the infinite refinement limit. This means in physical terms
that the TQFT has a trivial renormalization.

The continuum counterpart of the higher lattice gauge theory was
developed in~\cite{Ba02} and, for non-Abelian gerbes,
in~\cite{At02}. In the continuum, it has turned out to be difficult to
find fully gauge invariant actions. In the discrete approach, however,
see also~\cite{GrSc01}, invariant actions arise naturally from the
tetrahedron diagram, Figure~\ref{fig_commutative}. Our result is that
the admissible $2$-actions are sensitive only to $\ker t\unlhd H$, an
observation which may help to better understand the continuum
situation. In particular for the automorphism $2$-group, $\ker t$ is
often a discrete group. In this case, it will be impossible to write
down a naive continuum expression for the $2$-action which just uses
Lie algebras instead of Lie groups. Continuum $2$-actions will rather
involve some kind of topological defect or singularity and will be
given by non-local expressions similar to topological charges.

It should generally be possible to consider `infinitesimal' simplices
as in~\cite{At02} in order to infer the continuum formulas
corresponding to the given discrete expressions. It is open whether
the converse is possible, \ie\ to integrate the differential continuum
expressions in order to recover the non-infinitesimal formulas. The
problem is that there is no `surface ordered product' available
yet. Because of this obstacle, we favour the discrete approach, at
least for now.

Finally, the model constructed here together with the continuum
counterparts developed in~\cite{Ba02,At02} demonstrates that there
exist theories with local symmetries beyond conventional gauge
theory. Are the corresponding quantum field theories relevant in
nature? We have indicated two possible areas of physics in which they
might turn out to be useful, state sum models of quantum gravity and
the low energy behaviour of QCD.

\acknowledgments

I would like to thank Karl-Georg Schlesinger for discussions on the
construction of~\cite{GrSc01} in 1998 at HLRZ J\"ulich. It is a
pleasure to thank John Baez and Louis Crane for various conversations
on $2$-categories and $2$-groups. I am also grateful to Florian
Girelli, Robert Oeckl, Marni Sheppeard and Laurent Freidel for
discussions, comments on the literature or for carefully reading the
manuscript.

\end{document}